\documentclass[twocolumn,english,prx]{revtex4-1}
\usepackage{color}
\usepackage[utf8]{inputenc}
\usepackage{graphicx}
\usepackage[T1]{fontenc}
\usepackage{float}
\usepackage{textcomp}
\usepackage{amsmath}
\usepackage{amssymb}
\usepackage{graphicx}
\usepackage{esint}
\usepackage{natbib}
\usepackage{color}
\usepackage{ulem}

\begin{document}
\title{Electron spin inversion in fluorinated graphene nanoribbons}

\author{Bartłomiej Rzeszotarski}

\affiliation{AGH University of Science and Technology, Faculty of Physics and
Applied Computer Science,\\
 al. Mickiewicza 30, 30-059 Kraków, Poland}

\author{Alina Mre\'{n}ca-Kolasi\'{n}ska }

\affiliation{AGH University of Science and Technology, Faculty of Physics and
Applied Computer Science,\\
 al. Mickiewicza 30, 30-059 Kraków, Poland}

\author{Bartłomiej Szafran}

\affiliation{AGH University of Science and Technology, Faculty of Physics and
Applied Computer Science,\\
 al. Mickiewicza 30, 30-059 Kraków, Poland}

\begin{abstract}
We consider a dilute fluorinated graphene nanoribbon as a spin-active element.
The fluorine adatoms introduce a local spin-orbit Rashba interaction that induces spin-precession
for electron passing by. The spin precession involving a single adatom is infinitesimal
and accumulation of the 
 spin precession events with many electron passages under adatoms is necessary to accomplish a spin flip.
In order to arrange for this accumulation
 a circular n-p junction can be introduced to the ribbon by e.g. potential of the tip of an atomic force microscope. Alternatively  a
fluorinated quantum ring can be attached to the ribbon. 
We demonstrate that the spin-flip probability can be increased in this manner by as much as three orders of magnitude. 
The Zeeman interaction introduces spin-dependence of the Fermi wave vectors which changes the electron paths within the disordered system depending on the spin orientation.
The effect destroys the accumulation of the spin precession events in the n-p junction. For side attached quantum rings however - for which the electron path is determined
by the confinement within the channel -- the accumulation of the spin precession is robust against the Zeeman spin splitting.
\end{abstract}
\maketitle
\section{Introduction}

The coherent spin transport in semiconducting materials \cite{fabian} is under extensive studies
since the idea for the spin transistor was introduced \cite{datta}. Graphene \cite{castro} 
due to the absence of hyperfine interactions and  long spin coherence times \cite{han,maassen}
is a good candidate for a spin conductor. 
However, pristine graphene is a poor material for spin-active elements 
due to the weakness of the spin-orbit interaction \cite{manchon,laird}. Enhancement of the spin-orbit coupling
 was proposed by  deposition of graphene on transition metal dichalcogenides (TMDC) \cite{avsar,gmitra}.
The coupling with the TMDC layer gives rise to a strong Rashba interaction \cite{gmitra} that accompany redistribution
of the electron charge density 
and  an appearance of the electric field component perpendicular to the graphene plane. Alternatively, the spin-orbit interaction 
can be introduced by adatoms \cite{kotov}, hydrogen \cite{hy1,hy2,hy3,hy4,hy5,hy6,hy7} or fluorine \cite{fl1,fl2,fl3,fl4,fl5,fl6,irmer2015}.
The latter is ten times more efficient than the former as the source of the spin-orbit coupling \cite{irmer2015}.

In this paper we consider transport across the graphene nanoribbon \cite{waha} with a dilute fluorinated segment
as a spin inverter of the Fermi level electrons that could be implemented as a spin transistor channel \cite{datta}. 
The procedure to detect the spin flips on the electron motion within the spin-orbit coupled medium was recently demonstrated \cite{chuang}.
The system \cite{chuang} employs two quantum point contacts which are transparent to a single spin direction only. The rotation
of the electron spin on the way from one contact to the other results in the conductance drop. 

We study 
the effects  of the spin precession induced by the Rashba interaction near the adatoms.
We find that in the absence of the external magnetic field the spin-flip probability is very low, which we attribute
to cancellation of the spin precession effects by multiple electron backscattering since the direction of the spin precession in the Rashba field 
depends on the orientation of the electron momentum \cite{datta}.

In order to strengthen the spin-orbit coupling effects we introduce an n-p junction
in an external magnetic field. 
The n-p junctions in graphene can be induced by electrostatic potential  which moves the Dirac point above or below the Fermi level \cite{castro}. 
In the quantum Hall conditions these junctions form waveguides that confine currents \cite{wg1,wg2,wg3,wg5,wg4,wgx,wg11,wgk}.
The confinement in the classical terms can be understood as a result of the Lorentz force pushing the electrons to 
the n-p junction at both its sides.  The opposite orientation of the Lorentz force for the carriers of the conduction
and valence band in classical terms produces snake-orbits \cite{wg4,wgx,wg8,wg9,wg6,wg7,wg10,wg12,wgk} winding along the junction. 
The magnetic confinement of the current along the junction is supported for a single direction of the current only.
The effects of the spin-precession by the local Rashba interaction can be accumulated provided that the Fermi level electron 
passes many times under adatoms with  weak backscattering.
 In order to recycle the electron passages we consider a circular n-p junction defined by e.g. an external gate of the scanning probe microscope \cite{gorini,bhanda,tapa,mre1,mre2}.
In the quantum Hall conditions the current comes to the junction from the edge and the lifetime of the localized resonances with the current circulation
around the ring can be controlled by the gate potential, the Fermi energy \cite{mre1} and the external magnetic field \cite{mre2}.
We find that for the current circulating around the junction the spin-flip probability on the electron transfer can be increased by as much as three orders of magnitude.

We also consider the fluorinated ribbon in the quantum Hall conditions with the external potential removed. In strong magnetic fields
the localized resonances are associated with the current circulation from one adatom to the other and no backscattering along the path of incidence is possible. In these
conditions for a number of magnetic field values  the spin-flip probability produces high peaks. 
For low magnetic field we find a dip of conductance which is a signature of the weak localization effects    that in graphene are present  when the inter-valley scattering is strong \cite{wlg}.
In the present problem the adatoms as atomic scale perturbations to the lattice introduce strong intervalley scattering centers.

The effects of the accumulation of multiple spin flips requires that the electron trajectory remains unaffected by the spin orientation.
The dependence of the trajectory within the disordered system on the spin orientation appears via the Zeeman interaction and the resulting spin dependence of the Fermi wave vector.
In order to preserve large spin-flip probability the electron path across the fluorinated area needs to be weakly dependent on the wave vector. 
We show that this can be achieved with a fluorinated quantum ring side-attached to the ribbon. % for the values
For one of the perpendicular magnetic field orientations the Fermi level wave function is injected to the ring provided that the Fermi energy 
is in resonance with states circulating within the ring \cite{mre2}. The resulting spin precession is stable against the spin Zeeman interaction. Quantum rings are defined in graphene ribbons with a well established by etching  techniques \cite{qr1,qr2,qr3,qr4,szmel} and the role o the magnetic focusing has been discussed recently \cite{mre2,szmel}.

\section{Theory}
\subsection{Hamiltonian}
We use the atomistic $\pi$ band tight-binding Hamiltonian, which in the absence of the fluorine adatoms takes the form
\begin{equation}
H_0 =\sum_{i,\sigma} W_{i}c_{i,\sigma}^\dagger c_{i,\sigma}+ \sum_{\langle i,j \rangle}\sum_{\sigma}\left(t_{ij}c_{i,\sigma}^{\dagger}c_{j,\sigma}+H.c.\right)
\label{eq:ham-base}
\end{equation}
where $c_{i,\sigma}^{\dagger}$ and  $c_{i,\sigma}$
are the creation and annihilation operators for the electron on $i$-th ion with spin $\sigma$.  
The first sum introduces the external potential at the position of the $i$-th ion. The energy level 
corresponding to $p_z$ orbital is taken as the reference energy level. 
In the second summation $\langle i,j \rangle$ runs over nearest neighbor carbon atoms and $t=-2.7$ eV is the hopping parameter. 

\begin{figure}[htbp]
\centering
\includegraphics[width=0.29\textwidth]{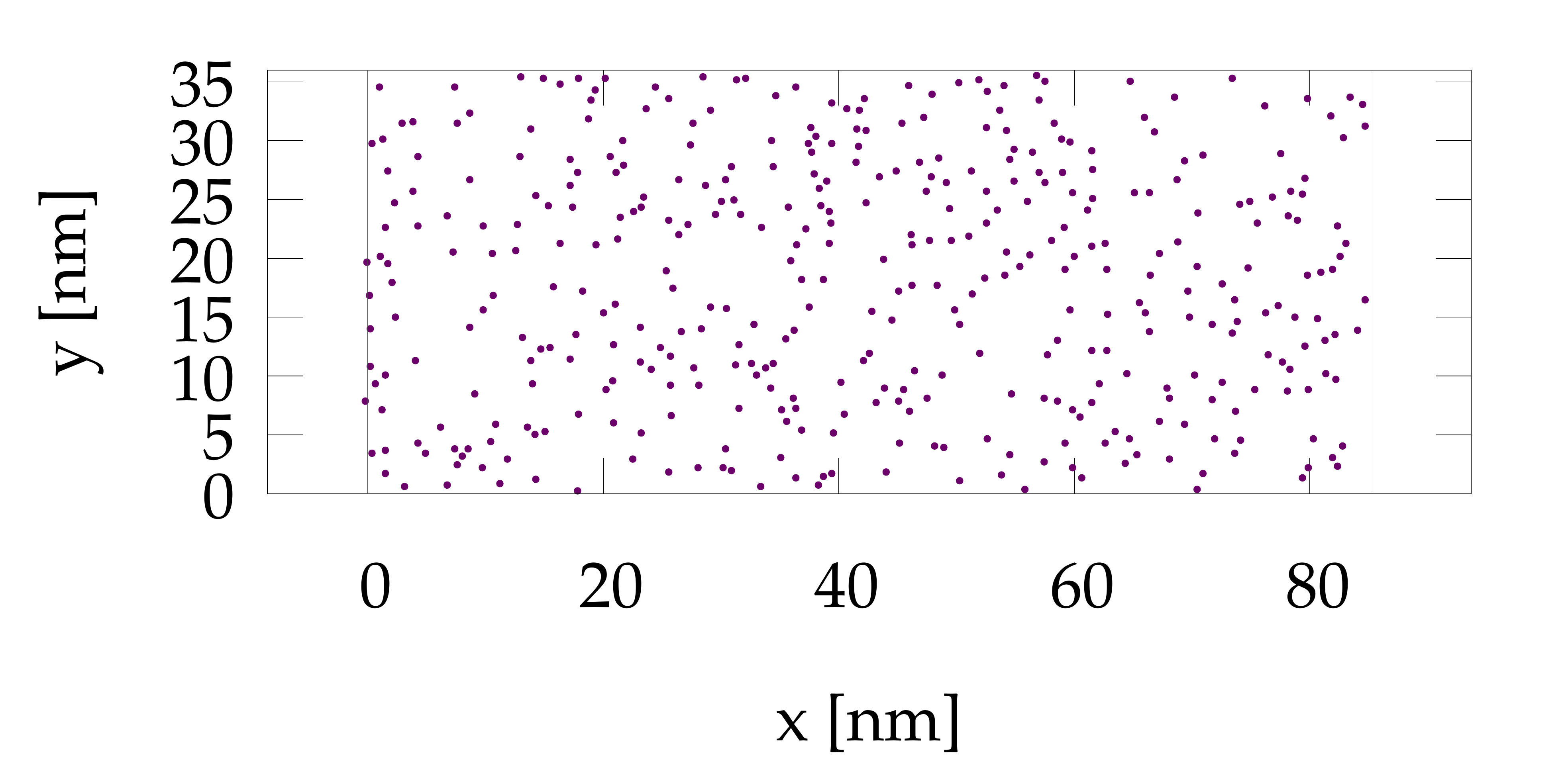}
\includegraphics[width=0.18\textwidth]{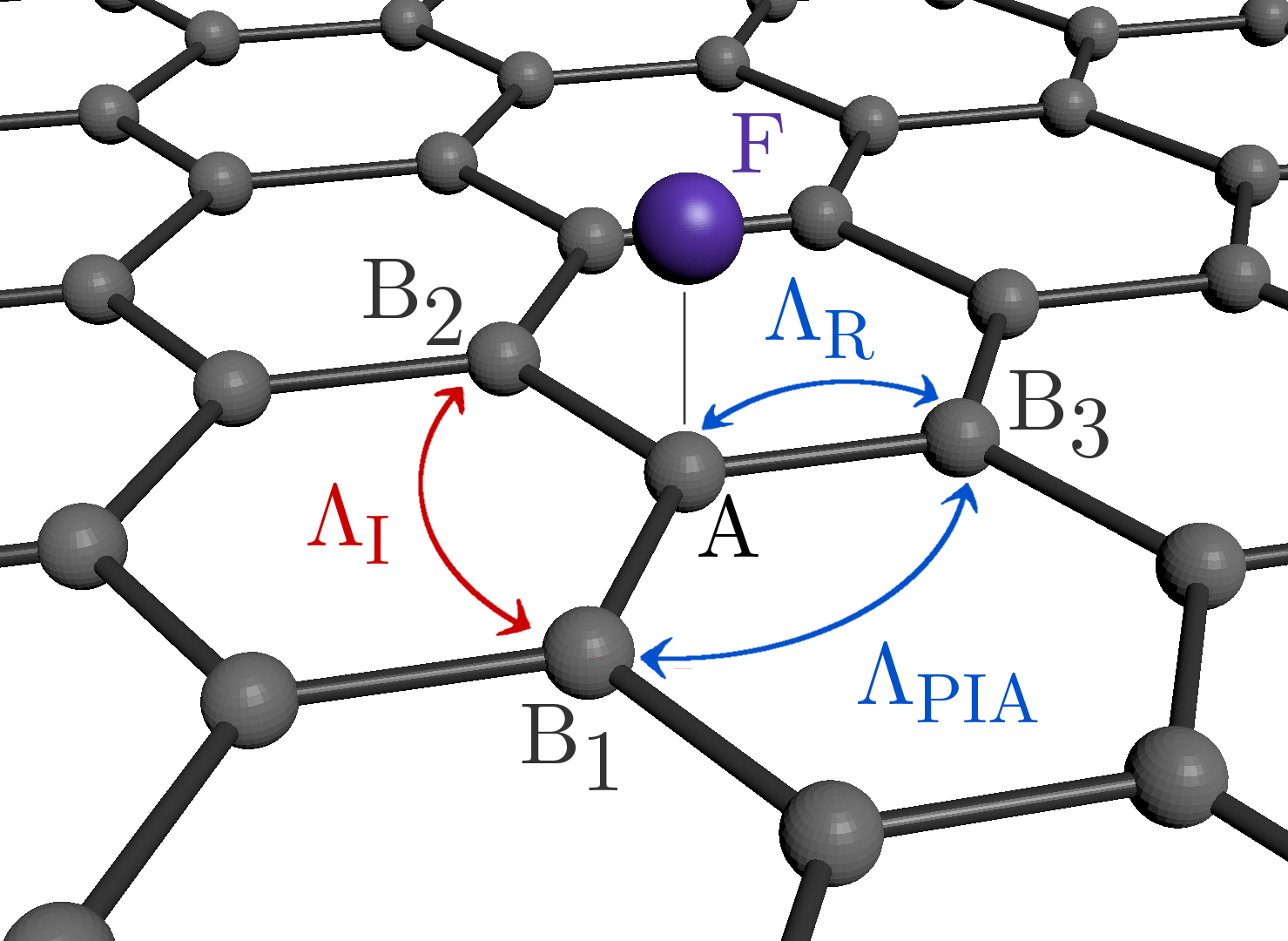}
\put(-242,65){(a)}
\put(-100,65){(b)}
\caption{ (a) The considered graphene nanoribbon with 292 atoms across and  the fluorinated segment. About 400 F adatoms are deposited at random (dots) within a part of nanoribbon 
that is $85.2\,$nm long. For the scattering problem the electron is incident from the left. The pristine graphene ends of the ribbon
are treated as the leads feeding and draining the charge and spin current from the system. (b) The fluorine adatom and its surroundings. The symbols refer to the local spin-orbit coupling Hamiltonian
introduced with the adatom $H_{SO}$ -- see Eq. (\ref{eq:ham-SO}).}
\label{fig:292-map}
\end{figure}

We consider a nanoribbon which in the absence of the adatoms is a strip of a crystal lattice of a finite width (see Fig.~\ref{fig:292-map}(a)).
The experiments on graphene ribbons indicate a presence of the transport gap \cite{hantg,chentg,evaldtg} near the charge neutrality point within which the system does not conduct electrical current.
For that reason  we consider here  a semiconducting armchair ribbon. 
 For the proof of principle in section III.A we take a ribbon that contains  only 3 carbon atoms on its width.
% ({\color{blue} tutaj: definiowanie "węższej" wstęgi niefortunnie może prowadzić do stwierdzenia, że jest węższa niż 3 atomy z poprzedniego zdania - malo istotne })
The rest of the results is obtained either for a thinner ribbon with 292 atoms across the width of $\simeq 35$ nm (subsections III.B, III.C, III.D, and III.F)
or for a wider ribbon (subsection III.E) with 1017 atoms across the width of $\simeq 125$ nm. 
For the thinner (wider) ribbon the scattering region considered in this work is a central finite region 85.2 nm (127 nm) long with fluorine 400 (2165) adatoms in a dilute concentration.
We also consider spin rotation within a fluorinated quantum ring of a circumference 283 nm [subsection III.F]. 
% ({\color{blue} XXX ile ich bylo w szerszej}).%  fluorine adatoms are present, which amounts in fluorination of the carbon atoms at the level of 0.5\%
%of carbon atoms fluorinated (see Fig. \ref{fig:292-map}) within the scattering region. %The ends of the computational box, where the boundary conditions are applied are free of adatoms.

The locations of the fluorine atoms are generated at random with a uniform distribution. 
The results presented below are typical and quantitatively independent of the specific distribution. The position of the conductance peaks
changes from one distribution to the other but the discussed physics does not. The fluorine atoms once adsorbed by graphene can 
only diffuse on its surface provided that the thermal excitations overcome migration barrier which for the fluorine adatoms is equal to 0.29 eV \cite{migratb}.
Therefore, the motion of the fluorine adatoms on the surface is frozen at low temperatures. %For higher temperatures the fluorine motion will be slow on the timescales of the electron dynamics.
The value changes with the carrier density level \cite{migratb2}, but below we consider mainly  conditions of the lowest subband transport, i.e. low Fermi energies near the charge neutrality point.

A fluorine adatom [see Fig.~\ref{fig:292-map}(b)] introduces additional terms to the Hamiltonian \cite{irmer2015},
\begin{equation}
H' = H_0 + H_{F} + H_{SO}, \label{totta}
\end{equation}
\noindent where $H_F$ stands for the orbital, and $H_{SO}$ the spin-orbit coupling effects.
The orbital contribution $H_F$ covers the level localized at the fluorine adatom ($F$) with the energy level $\varepsilon_f=-2.2$ eV 
and the hopping element $T=5.5$ eV between the fluorine and the carbon atom ($A$) that form a vertical dimer ($F-A$, see Fig.~\ref{fig:292-map}(b)):
\begin{equation}
H_F= \varepsilon_f \sum_{\sigma} \hat{F}^\dagger_{\sigma} \hat{F}_{\sigma} + T \sum_{\sigma}( \hat{F}^\dagger_{\sigma} \hat{A}_{\sigma} + \hat{A}^\dagger_{\sigma} \hat{F}_{\sigma}),
\label{eq:ham-flu}
\end{equation}
where $\hat{F}_{\sigma}$ is the annihilation  operator for electron in the fluorine ion with spin $\sigma$, 
and $A_\sigma$ is the annihilation operator for the fluorinated carbon atom. 

The three next nearest neighbor atoms of the fluorinated carbon $A$ are denoted by $B$ (see Fig.~\ref{fig:292-map}).
 The annihilation operator for the electron in the $B$ atoms is $\hat{B}^\dagger_{i,\sigma}$.
With this notation the spin-orbital part has the form \cite{irmer2015}
\begin{align}
H_{SO} &= \frac{i\Lambda^\text{B}_\text{I}}{3\sqrt{3}}\sum_{\mathbf{\langle}\langle i,j \rangle \rangle} \sum_{\sigma} \hat{B}^\dagger_{i,\sigma}\nu_{ij}(\hat{{\bf s}}_z)_{\sigma\sigma}\hat{B}_{j,\sigma} \nonumber \\
&+\frac{2i\Lambda_\text{R}}{3}\sum_{B_j \in C_{nn}} \sum_{\sigma\neq\sigma'} [\hat{A}^\dagger_{\sigma}(\hat{\mathbf{s}}\times\mathbf{d}_{\text{A}j})_{z,\sigma\sigma'}\hat{B}_{j,\sigma'}+\text{H.c.}] \nonumber \\
&+\frac{2i\Lambda^\text{B}_\text{PIA}}{3}\sum_{\mathbf{\langle}\langle i,j \rangle \rangle} \sum_{\sigma\neq\sigma'} \hat{B}^\dagger_{i,\sigma}(\hat{\mathbf{s}}\times\mathbf{d}_{ji})_{z,\sigma\sigma'}\hat{B}_{j,\sigma'}.
\label{eq:ham-SO}
\end{align}
The first sum in the intrinsic spin-orbit coupling in the Kane-Mele form \cite{km} which is spin-diagonal. The second term describes the Rashba interaction induced by the perpendicular electric field due to the deformation
of the electron density by the adatom. The last term is the pseudospin-inversion-asymmetry next-nearest neighbor term that results from deformation of the graphene lattice by the adatom \cite{liu}.  The summation 
 $\langle\langle i,j \rangle\rangle$ runs over the  nearest neighbors of the fluorinated atom $\{\text{B}_1,\text{B}_2,\text{B}_3\}$.  The coefficient 
$\nu_{ij}$ is $+1$ ($-1$) when the path from $j$ to $i$ via a common nearest neighbor $k$, $j \rightarrow k \rightarrow i$, is counterclockwise (clockwise)
and ${\mathbf d}_{ij}$ is the unit vector in the $xy$ plane oriented from ion $j$ to $i$.
The applied spin-orbit coupling parameters \cite{irmer2015} for $0.5\%$ concentration of the fluorine atoms are $\Lambda^\text{B}_\text{I}=3.3\,$meV,  $\Lambda_\text{R}=11.2\,$meV and $\Lambda^\text{B}_\text{PIA}=7.3\,$meV. Both the Rashba and the PIA terms induce spin variation in the electron motion across the fluorinated area. However, for the applied parameters the effect of the PIA is by an order of magnitude lower 
in terms of the spin-flipping transfer probability.

The orbital effects of the external perpendicular external magnetic field is introduced to the Hamiltonian by modification of the hopping parameters. 
For the Hamiltonian of Eq. (\ref{totta}) put in a general form,
\begin{eqnarray} 
H'=\sum_{k,l,\sigma,\sigma'} h_{k\sigma l\sigma'}c_{k\sigma}^\dagger c_{l\sigma'}, 
\end{eqnarray}
in presence of the external magnetic field the hopping parameters are modified by the Peierls phase
\begin{eqnarray} 
H'_B&=&\sum_{k,l,\sigma,\sigma'} h_{k\sigma l \sigma'} \exp\Big[ \frac{2\pi i}{\Phi_0} \int^{\mathbf{r}_j}_{\mathbf{r}_i} \mathbf{A}\cdot \text{\textbf{dl}} \Big]c^\dagger_{k\sigma}c_{l\sigma'} %+e F_z \sum_{k,\alpha} z_k c^\dagger_{k,\alpha}c_{l,\alpha},  
\label{genio}
\end{eqnarray}
where $\mathbf{A}$ is the vector potential,  $\Phi_0=h/e$ is the magnetic flux quantum and $\mathbf{r}_i$ is the position of $i$-th ion.

The spin-effects of the magnetic field are introduced by the Zeeman interaction
\begin{eqnarray}
H'_{B,Z}=H'_B+ \frac{1}{2}g\mu_B B\sum_{k,\sigma}  \left(\hat{\bf s}_z\right)_{\sigma\sigma} c_{k\sigma}^\dagger c_{l\sigma},
\end{eqnarray}
 with $\mu_B$  the Bohr magneton and $g=2$.

\subsection{Quantum rings: induced at the n-p junction and side-attached}

\begin{figure}[htbp]
\centering
\includegraphics[width=0.4\textwidth]{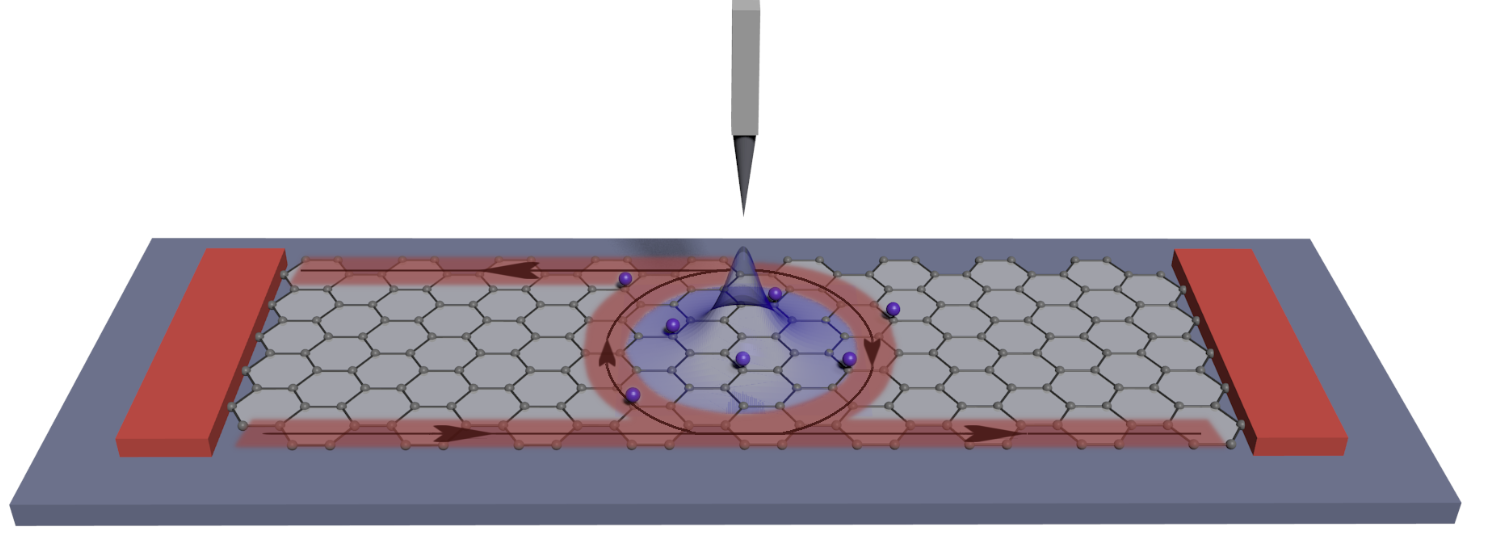}
\caption{The nanoribbon with $F$ adatoms, and the n-p junction introduced by a tip of an atomic force microscope. The p region is induced below the tip. The arrows show
the orientation of the edge and junction currents for the magnetic field oriented from above to the graphene plane.  }\label{s2}
\end{figure}

In order to form an n-p junction within the ribbon we introduce an external potential of e.g. a charged tip of an atomic force microscope \cite{mre1,mre2,kolasinski2013} -- see Fig.~\ref{s2}.
The original tip potential is of the Coulomb form which is screened by deformation of the electron gas within the conducting plane.
The Schr\"odinger-Poisson equations for the tip screened by the two-dimensional electron gas produce an effective tip potential \cite{kolasinski2013} 
which is close to the Lorentz form 
\begin{equation}
W(x,y) = \frac{V_t}{1+[(x-x_t)^2+(y-y_t)^2]/d^2},
\label{eq:lorentz}
\end{equation}
where $x_t,y_t$ is the tip position over the plane, and $d$ is the effective width of the tip potential that is of the order of the tip-electron gas distance \cite{kolasinski2013}), and  $V_t$ depends on the charge accumulated by the tip. We adopt $V_t=400$ meV and $d=4.92$ nm as in the previous paper \cite{mre1}.
In Hamiltonian $H_0$ given by Eq. (1), $W_i=W(x_i,y_i)$. For the workpoint we set $E_F=60$ meV with the fluorine concentration $\eta = 0.5\%$  unless stated otherwise.

\begin{figure}[htbp]
\centering
\includegraphics[width=0.4\textwidth]{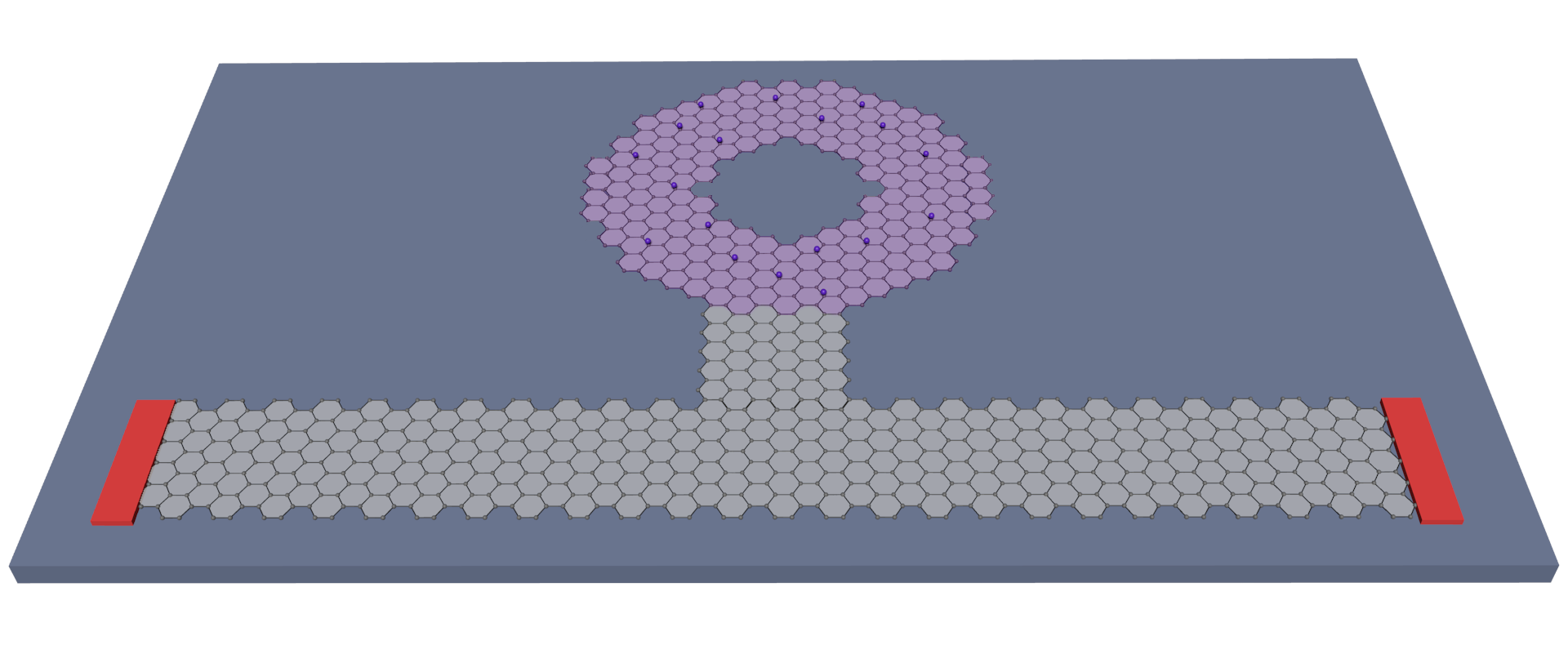}
\caption{Schematics of a fluorinated quantum ring side-attached to the graphene ribbon.}\label{ring}
\end{figure}

In presence of the Zeeman effect large spin-flips can be obtained when the electron path remains the same for both spin orientations. We arange
for these conditions with a quantum ring side attached to the ribbon -- see Fig.~\ref{ring}. The ribbon that is 293 atoms wide is considered for this purpose. 
The external and internal radii of the ring are 27 nm and 63 nm, respectively, and the arm of the ring contains about 293 atoms along its radius. The fluorine is
adsorbed only within the ring.

\subsection{Solution of the scattering problem}

\subsubsection{Implementation of the Landauer approach}

Outside the scattering region the nanoribbon does not contain adatoms, so that the input and the output leads, i.e.,
the left and right ends of the ribbon, respectively, are free of the spin-orbit interaction. 
The eigenstates of the leads are spin-independent and are determined in the Bloch form 
\begin{equation}
\psi_{u,v,\sigma}^{k_m}=\chi^{k_m}_{v,\sigma}e^{ik_mu\Delta x},
\label{eq:bloch}
\end{equation}
where $k_m$ is the wave vector for  $m$-th subband, $u$ numbers the elementary cells of the ribbons, and $v$ numbers the atoms inside the elementary cell.
The wave function in the input lead is a superposition of the incident electron states with wave vectors $k^+$ and backscattered states with wave vectors $k^-$,
\begin{equation}
\Psi_{u,v,\sigma}^{in} = \displaystyle\sum_l \Big( c_{in}^l\chi_{v,\sigma}^{k_l^+}e^{ik_l^+u\Delta x} + d_{in}^l\chi_{v,\sigma}^{k_l^-}e^{ik_l^-u\Delta x} \Big),
\label{eq:psi-in}
\end{equation}
where the summation over $l$ runs over the subbands appearing at the Fermi level. 
At the output channel we have only the transferred, right-going waves 
\begin{equation}
\Psi_{u,v,\sigma}^{out} = \displaystyle\sum_l c_{out}^l\chi_{v,\sigma}^{k_l^+}e^{ik_l^+u\Delta x}.
\label{eq:psi-out}
\end{equation}
 We solve for the scattering wave function $\Psi$ that is an eigenstate
of the Hamiltonian for an assumed Fermi energy that is glued to the boundary conditions given by Eqs. (\ref{eq:psi-in}) and  (\ref{eq:psi-out}) for each incoming subband separately setting $c_{in}^l = \delta_{lm}$, for the electron incident from the subband $m$.
We use the wave function matching method \cite{wfmm}, which produces 
the scattering amplitudes for the reflected  ($d_{in}$) and transferred ($c_{out}$) waves.

For the Hamiltonian in form (\ref{genio}) the current flowing along the $\pi$ bonds between $m$ and $n$ ions  is calculated as 
\begin{equation}
J_{k\sigma l\sigma'}=\frac{i}{h}\sum_{\sigma,\sigma'}[h_{k\sigma l\sigma'}\Psi^*_{k,\sigma}\Psi_{l,\sigma'}-h_{l\sigma'k\sigma}\Psi^*_{l,\sigma'}\Psi_{k,\sigma}],
\label{eq:curr}
\end{equation}
where $\Psi_{n,\sigma}$ is the $\sigma$ spin component of the wave function at the $n$-th ion.
The formula (\ref{eq:curr}) accounts for the transfer of the current from one component of the spin to the other provided that the Hamiltonian contains the spin-orbit coupling components. 
This is indeed the case near the adatoms. Outside the fluorinated area the spin currents are conserved and can be calculated from Eq (\ref{eq:curr}) for each component of $\sigma_z$.  

We calculate the current fluxes outside the fluorinated area, at the ends of the ribbon which serve as the input and output leads,
\begin{equation}
\phi = \sum_{  \sigma m,n_m} J_{m\sigma n_m \sigma},
\label{eq:flux}
\end{equation}
where $m$ are the indices of the atoms across the ribbon and $n_m$ are their left (right) nearest neighbors in the ribbon left (right) lead where the flux is calculated. 
The electron transfer probability from the subband $n$ to the subband $m$  is calculated as 
\begin{align}
T_{mn} &= \Big\vert \frac{c^m_{out}}{c^n_{in}}\Big\vert^2 \Big\vert \frac{\phi_{k^+_m}}{\phi_{k^+_n}}\Big\vert,  \\
\end{align}
where $c_{out}^m$ and $c^n_{in}$ are the scattering amplitudes of the transferred wave function and incident wave functions,
in the $m$ and $n$ modes, respectively.
The overall transfer probability from the $n$-th subband is $T_n = \sum_m T_{mn}$,
which is used in the Landauer
formula for conductance
\begin{equation}
G = G_0 \sum_n T_n, \end{equation}
where $G_0=e^2/h$ is the flux quantum. 
In the following we set the incident electron spins and the Bloch functions in the leads as eigenstates of the $\sigma_z$ operator.
We calculate the spin conserving $G_{uu}$, $G_{dd}$, and spin flipping conductance contributions
$G_{ud}$ and $G_{du}$ by summation of $T_{mn}$ with respect to the two spin orientations, where $d$ and $u$ stand for the $\sigma_z$
eigenstates with eigenvalues -1 and +1, respectively.

\subsubsection{Boundary conditions}
\label{ss:boundary}
Boundary conditions for left (input) and right (output) lead was calculated using Wave Function Matching (WFM) method \cite{wfmm}.
The Hamiltonian matrix for infinite ribbon that serves as the input or output lead to the scattering region can be put in form

\begin{equation}
\mathcal{H}=
\begin{pmatrix}
 \ddots & \cdots & 0 & 0 & \   \\
 \vdots & \mathbf{H}_{u-1} & \mathbf{B}^{\dagger}_{u-1} & 0 & 0 \\
 0 & \mathbf{B}_{u-1} & \mathbf{H}_{u} & \mathbf{B}^{\dagger}_{u} & 0 \\
 0 & 0 & \mathbf{B}_{u} & \mathbf{H}_{u+1} & \vdots \\
  & 0 & 0 & \cdots & \ddots
\end{pmatrix},
\end{equation}

\noindent where matrix $\mathbf{H}_{u}$ is the Hamiltonian of single elementary cell $u$, matrix $\mathbf{B}_{u}$ describes connections between cells $u$ and $u+1$. Matrices $\mathbf{H}_{u}$ and $\mathbf{B}_{u}$ are the size of $2n \times 2n$ where $n$ is the orbital number for the atom inside elementary cell (factor 2 arise from spinor splitting). For an ideal periodic infinite ribbon matrices $\mathbf{H}_{u}$ and $\mathbf{B}_{u}$ are identical, therefore the $u$ symbol will be omitted from this point. The eigenfunction of the Hamiltonian $\mathcal{H}$ can be divided on eigenfunctions for each elementary cell 
\begin{equation}
\boldsymbol{\psi}=
\begin{pmatrix}
\vdots \\
\psi_{u-1} \\
\psi_{u} \\
\psi_{u+1} \\
\vdots
\end{pmatrix}
\end{equation}

\noindent where $\psi_{u}$ is the vector of size $2n$ ($n$ orbitals for each spinor). The wave function satisfies the equation

\begin{equation}
-\mathbf{B}\psi_{u-1} +(E\mathbf{I}-\mathbf{H})\psi_{u} - \mathbf{B^{\dagger}}\psi_{u+1} = 0.
\end{equation}

\noindent Using Bloch form of the wave function (\ref{eq:bloch}) in substitution

\begin{equation}
\psi_{u-1} = \chi, \;\;\; \psi_{u}=\lambda\chi , \;\;\; \psi_{u+1}=\lambda^2\chi, \nonumber
\end{equation}

\noindent where $\lambda=\exp(ik\Delta x)$, we get

\begin{equation}
-\mathbf{B}\chi +\lambda(E\mathbf{I}-\mathbf{H})\chi - \lambda^2\mathbf{B^{\dagger}}\chi = 0,
\label{eq:abvwfm}
\end{equation}

\noindent For  $\eta = \lambda \chi$ equation (\ref{eq:abvwfm}) above can be written as

\begin{equation}
\begin{bmatrix}
\begin{pmatrix}
0 & \mathbf{I} \\
-\mathbf{B} & E\mathbf{I}-\mathbf{H}
\end{pmatrix}
-\lambda
\begin{pmatrix}
\mathbf{I} & 0 \\ 
0 & \mathbf{B^\dagger}
\end{pmatrix}
\end{bmatrix}
\begin{bmatrix}
\chi \\ \eta
\end{bmatrix}
=0.
\label{eq:bound-eig}
\end{equation}

\noindent Described eigenproblem has $4n$ solutions: $2n$ left-going and $2n$ right-going modes - decaying or propagating. It is straightforward to identify right and left going evanescent modes. The eigenvalue satisfies $\vert \lambda_{n,\sigma} \vert < 1$ for right-going evanescent modes and $\vert \lambda_{n,\sigma} \vert > 1$ for left-going evanescent modes. For the propagating modes of Bloch waves the eigenvalue satisfies $\lambda_{n, \sigma}=\exp(ik\Delta x)$, with real $k$, hence $\vert \lambda_{n,\sigma} \vert = 1$. Then we look for values of $\vert \lambda \vert=1$ for a given $E$ and determine the corresponding wave vectors and periodic functions $\chi$. 

\subsubsection{Calculation of the scattering amplitudes}
Transmission and reflection probabilities  are calculated using scattering coefficients  $c_{out}^l$ and $d^l_{in}$, respectively. 
The scattering wave function in the left output lead takes the form

\begin{eqnarray}
&&\Psi_{0,v,\sigma}^{in} \nonumber  = c_{in}\chi_{v,\sigma}^{k_{in}^+} \nonumber +\sum_l d_{in}^l\chi_{v,\sigma}^{k_{l}^-},
\label{eq:psi-in-trans}
\end{eqnarray}
where the term with $c_{in}$ describes the incident electron and the sum stands for the superposition of the backscattered electron waves. 
\noindent We calculate the derivative of the wave function 

\begin{align}
\frac{1}{\Delta x}(\Psi_{0,v,\sigma}^{in} - \Psi_{-1,v,\sigma}^{in}) = 
\sum_l c_l\chi_{v,\sigma}^{k_{l}^+}\frac{1}{\Delta x}(1-e^{-ik_{l}^+\Delta x})+& \nonumber \\\sum_l d_{in}^l\chi_{v,\sigma}^{k_{l}^-}\frac{1}{\Delta x}(1-e^{-ik_{l}^-\Delta x}),
\end{align}
\noindent from where we get the equation for the wave function outside the computational box 

\begin{equation}
\Psi_{-1,v,\sigma}^{in} = \Psi_{0,v,\sigma}^{in} - \sum_l c_l\chi_{v,\sigma}^{k_{l}^+}\Delta_{k_{l}^+} -\sum_l d_{in}^l\chi_{v,\sigma}^{k_{l}^-}\Delta_{k_{l}^-}.
\end{equation}

Using scalar product for the wave function in the first cell and the function $\chi^{k_{l'}}$ we obtain

\begin{equation}
\langle \chi^{k_{l'}^-}_{v,\sigma}\vert\Psi_{0,v,\sigma}^{in}\rangle = \sum_l c_l \langle \chi^{k_{l'}^-}\vert\chi^{k_l^+}\rangle + \sum_l d^l_{in} \langle \chi^{k_{l'}^-}\vert\chi^{k_l^-}\rangle,
\label{eq:left-bound-out}
\end{equation}

\noindent where $\langle A \vert B \rangle = \displaystyle\sum_v A_v^* B_v$ denote the inner product in discrete form. Defining matrices $\mathbf{B}_{l',l}=\langle \chi^{k_{l'}^-} \vert	\chi^{k_{l}^+}\rangle$, $\mathbf{S}_{l',l}=\langle \chi^{k_{l'}^-} \vert	\chi^{k_{l}^-} \rangle$ and the vector $\mathbf{A}_{l'}=\langle \chi^{k_{l'}^-} \vert	\Psi_0^{in} \rangle$ we can rewrite Eq. (\ref{eq:left-bound-out}) in a matrix form

\begin{equation}
\mathbf{A} = \mathbf{Bc}_{in} + \mathbf{Sd}_{in},
\end{equation}

\noindent which satisfies

\begin{equation}
d^l_{in} = \sum_{l'}(\mathbf{S}^{-1})_{l,l'}\mathbf{A}_{l'} - \sum_{l',j}(\mathbf{S}^{-1})_{l,j}\mathbf{B}_{j,l'}\mathbf{c}^{l'}_{in} .
\label{eq:din}
\end{equation}

\noindent Using Eq. (\ref{eq:din}) in Eq. (\ref{eq:left-bound-out}) we obtain the left boundary condition.
For the right end ($N$ cell) the wave function
\begin{equation}
\Psi_{N,v,\sigma}^{out} = \sum_l c_{out} \chi_{v,\sigma}^{k_l^+}e^{-iNk_l^-\Delta x}.
\label{eq:psi-out-trans}
\end{equation}

\noindent Using similar approach to the right end of the computational box

\begin{align}
\frac{1}{\Delta x}(\Psi_{N+1,v,\sigma}^{out} - \Psi_{N,v,\sigma}^{out}) &= \sum_l c_{out}^l\chi_{v,\sigma}^{k_{l}^+}\frac{1}{\Delta x}e^{iNk_{l}^+\Delta x}(e^{ik_{l}^-\Delta x}-1)\nonumber \\&=\sum_l c_{out}^l\chi_{v,\sigma}^{k_{l}^+}\frac{1}{\Delta x}e^{iNk_{l}^+\Delta x}\Delta'_{k_l^+},
\end{align}
\noindent from which we get the wave function outside the right side of the computational box

\begin{equation}
\Psi_{N+1,v,\sigma}^{out} = \Psi_{N,v,\sigma}^{out} + \sum_l c_l\chi_{v,\sigma}^{k_{l}^+}e^{iNk_{l}^+\Delta x}\Delta'_{k_l^+}.
\end{equation}

\noindent Using scalar product 

\begin{align}
\langle \chi^{k_{l'}^+}_{v,\sigma}e^{iNk_{l'}^+\Delta x}\vert\Psi_{N,v,\sigma}^{out}\rangle \nonumber & = \sum_l c^l_{out} \langle \chi^{k_{l'}^+}e^{iNk_{l'}^+\Delta x}\vert\chi^{k_l^+}e^{-iNk_{l'}^+\Delta x}\rangle \\ &= \sum_l \mathbf{S'}_{l',l}c^l_{out},
\end{align}
\noindent we get
\begin{equation}
c^l_{out} = (\mathbf{S'}_{l',l})^{-1} \langle \chi^{k_{l'}^+}_{v,\sigma}e^{iNk_{l'}^+\Delta x}\vert\Psi_{N,v,\sigma}^{out}\rangle.
\end{equation}

\section{Results and Discussion}
\subsection{Accumulation of the spin precession events $(B=0)$}
As a proof of principle for accumulation of the local spin precession evens we consider a narrow ribbon depicted in Fig.~\ref{pp}. 

The ribbon is a sequence of hexagons with the fluorine atoms placed near the junctions of one hexagon to the other [Fig. \ref{pp}]. No external
magnetic field is applied and the electron is injected to the system with spin-down orientation from the left-hand side. 
The positions of the F atoms are repeated periodically within the ribbon, so the $G$ dependence on $E_F$ forms
a series of resonances as for a superlattice band filter \cite{pacher}. 
 We set the value of the Fermi energy $E_F=4.620512$ eV marked by the dot in Fig.~\ref{trans} for which the 
system is transparent for electrons and look at the average spin $z$ and $x$ components calculated for subsequent hexagonal elementary cells of the ribbon 
in Fig.~\ref{spop}.  
For the resonant energy  electron on its way across the system comes to fluorinated carbon atom with the same momentum and no backscattering is present.
 In total there are 400 $F$ 
atoms along the ribbon with the concentration $\eta \approx 6.7\%$. 
The overall conductance $G$ and the spin-flip contribution are depicted in Fig.~\ref{trans}. 

Although an effect of a single fluorine atom to the orientation of the electron spin is very small we can see that upon transition below 400 fluorine atoms within the system which is nearly enough to rotate the electron spin {from the $-z$ the $+z$ orientation. }
The source of the spin-flip transfer in the considered system is the Rashba component of the Hamiltonian
that is due to the local electric field introduced by adatoms. 

The Rashba spin-orbit interaction due to the  adatom introduces an effective magnetic field \cite{meier} ${\bf B}_{SO}=C\left({\bf p}\times {\bf E}\right)$ \cite{meier},
where $C$ is a  constant, ${\bf p}$ the electron momentum and ${\bf E}$ the electric field vector. For the planar motion $p_z=0$ 
and the vertical electric field $E_z\neq 0$ induced by the adatom, the ${\bf B}_{SO}$ is oriented along the $y$ direction, i.e. in-plane and perpendicular
to the momentum orientation. The spin of the electron moving within the graphene layer precesses around ${\bf B}_{SO}$ \cite{datta,chuang,besza} -- in the case
considered in Fig.~\ref{spop} from the orientation to below the ribbon ($-z$) {through the orientation along the ribbon ($x$) for about 480th hexagonal cell of the ribbon  to the orientation above the ribbon ($+z$) for the 960th cell.

\begin{figure}[htbp]
\includegraphics[width=0.42\textwidth]{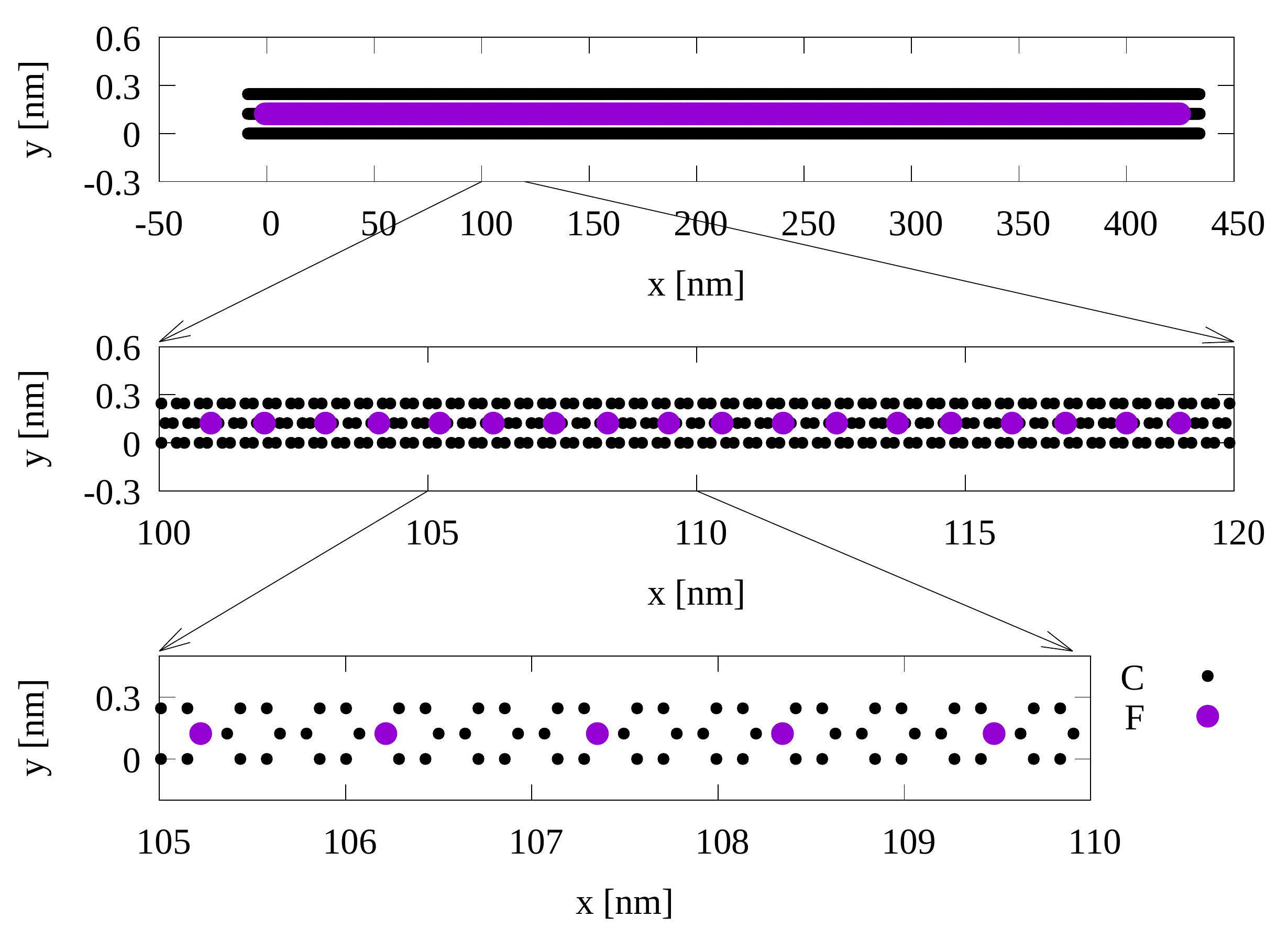}
\caption{A narrow ribbon that is considered in subsection III.A with the fluorine concentration  $\eta \approx 6.7\%$. The position of the C and F ions are marked with black and purple dots, respectively.}
 \label{pp}
\end{figure}

\begin{figure}[htbp]
\includegraphics[width=0.5\textwidth]{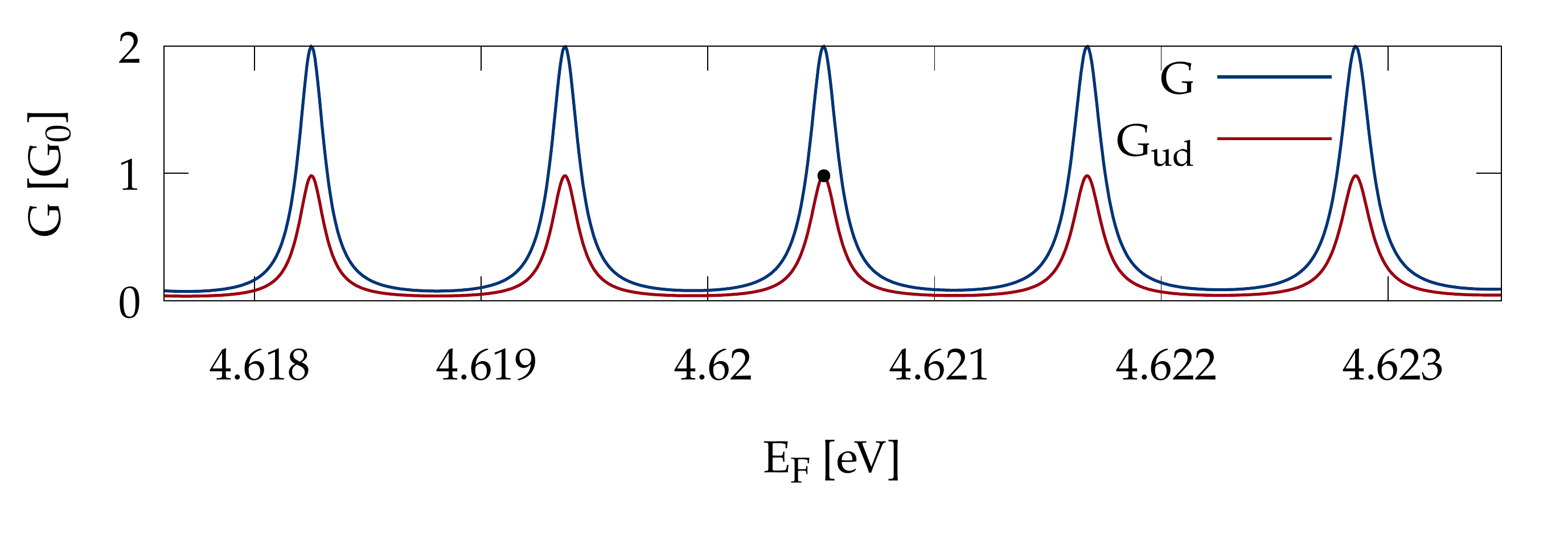}
\caption{The conductance $G$ and its spin-flip contributions for the system given in Fig.~\ref{pp}. The narrow range of $E_F$ was chosen to distinguish resonances.
%{\bf p. Bartek: nie mozna dawac opisow krzywych poza osiami. PRB nie przyjmuje takiego formatu. takze ograniczenie na szerokosc rysunku. $E_F$ zamiast $E$}
}
 \label{trans}
\end{figure}

\begin{figure}[htbp]
\includegraphics[width=0.5\textwidth]{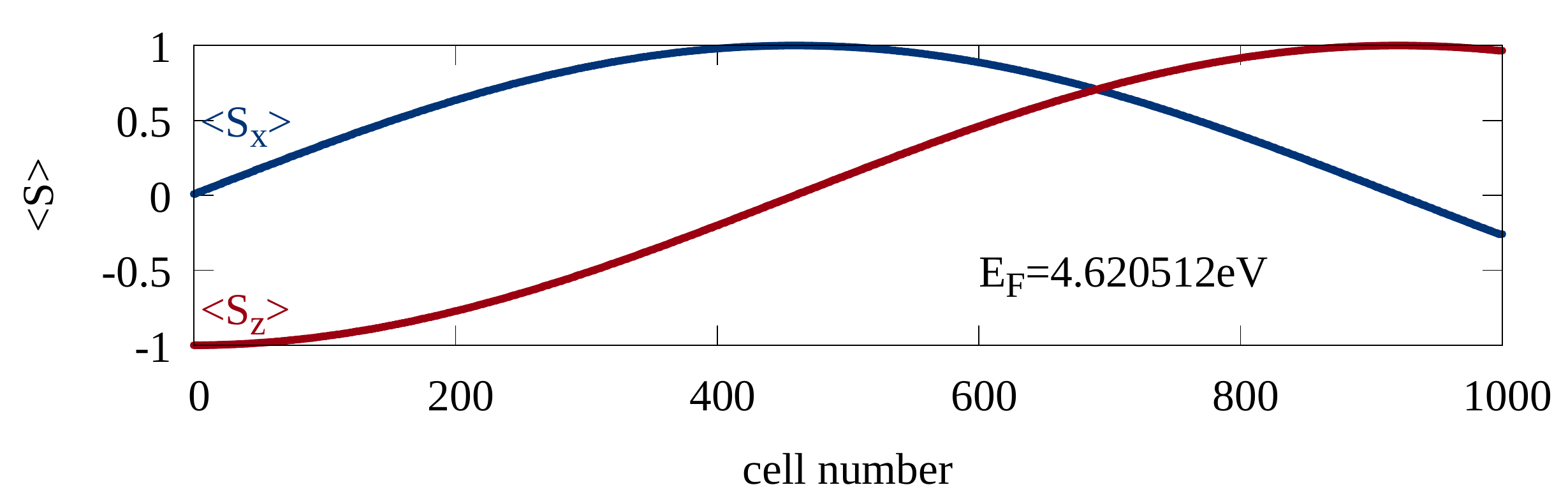}
\caption{The average $x$ and $z$ components of the spin calculated for subsequent elementary cells for $E_F=4.620512$ eV (see the dot in Fig.~\ref{trans}) 
for which the system is transparent for the Fermi electron. The electron is incident from the left with ${\bf s}_z=-1$ orientation. The average $y$ component (not shown) is nearly 0 all along the ribbon.  
%{\bf p. Bartek: nie ma potrzeby dwoch paneli, prosze jeden zrobic. Os pozioma od 0 do 4000.}
}
 \label{spop}
\end{figure}

\begin{figure}[htbp]
\centering
\begin{tabular}{ll}
(a)& \\ (b) & \includegraphics[width=0.42\textwidth]{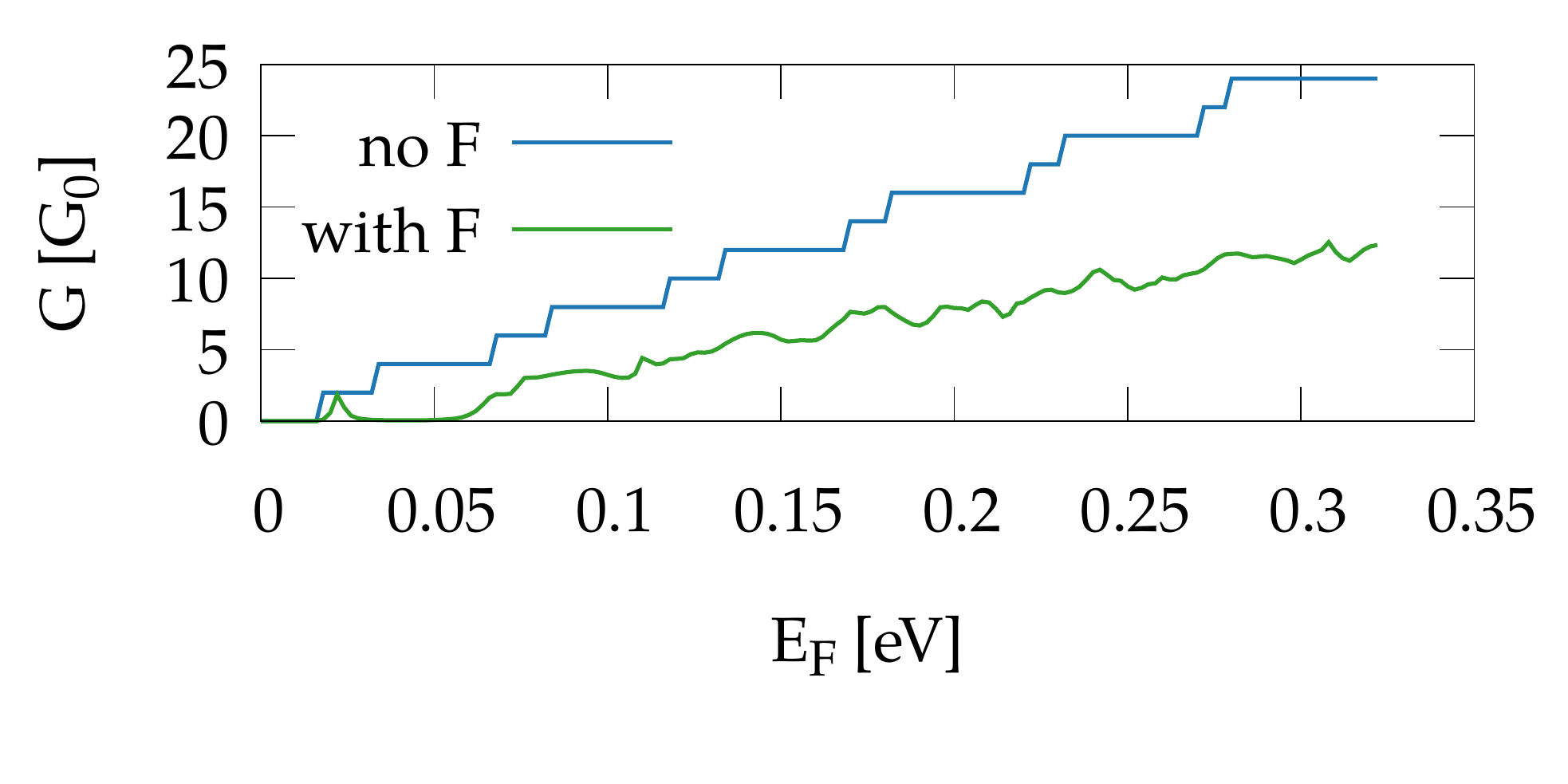}\\
(c) &\includegraphics[width=0.42\textwidth]{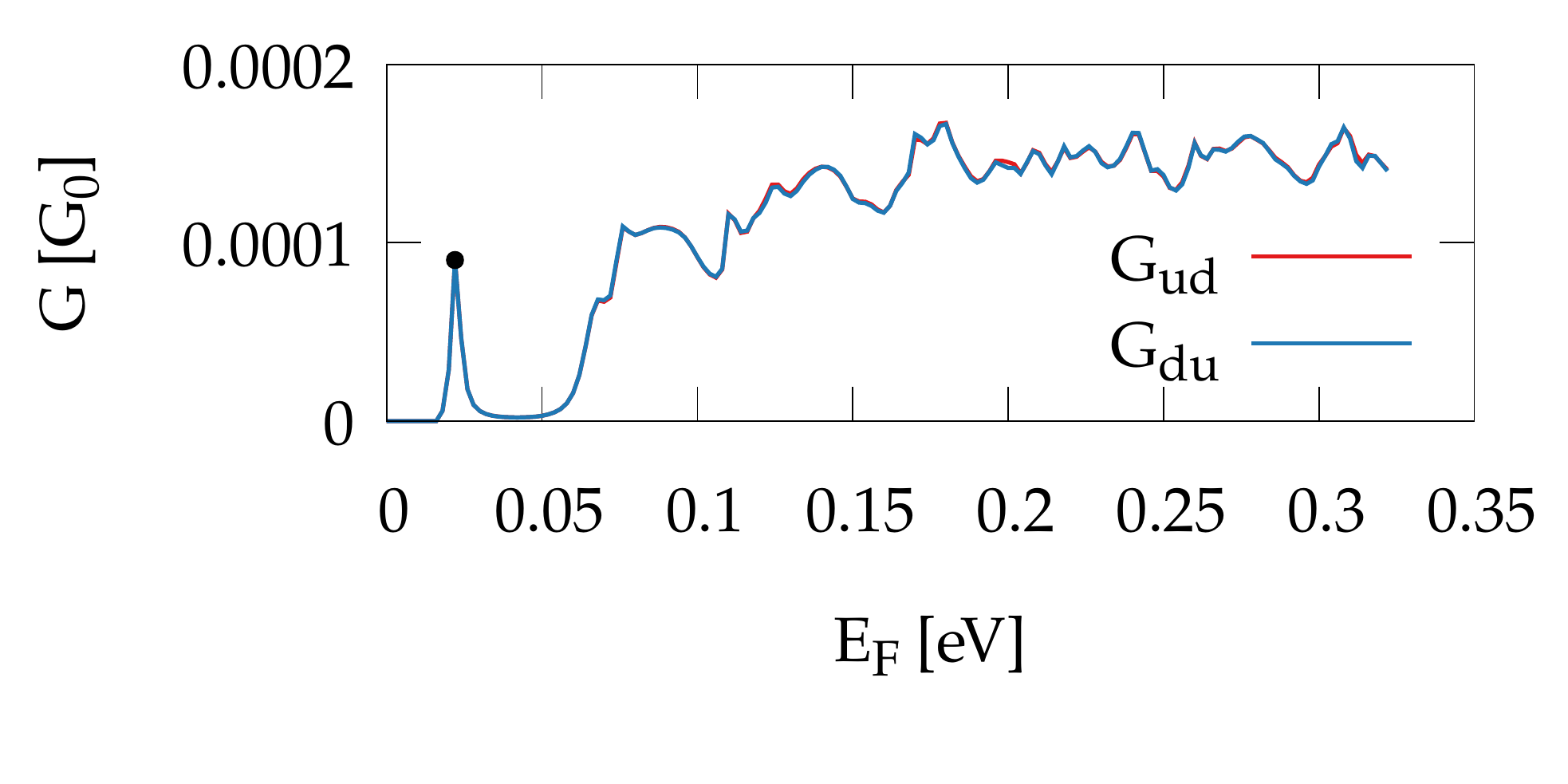}\\
 &\includegraphics[width=0.39\textwidth]{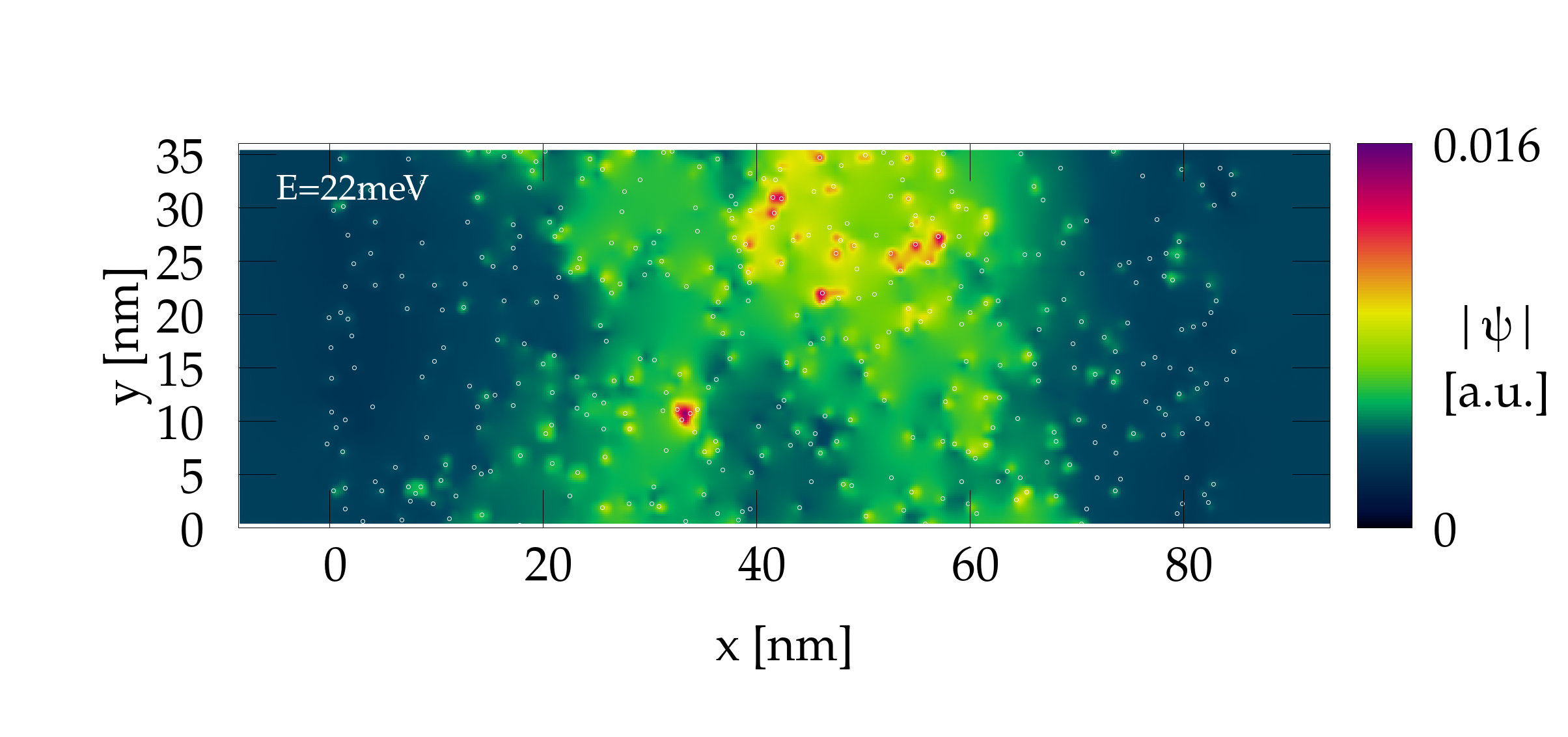}
\end{tabular}
\caption{ (a) Conductance  for a pristine semiconducting armchair nanoribbon (blue line) and 
the one with 400 fluorine adatoms (green line) as functions of the Fermi energy at $B=0$. 
(b) The spin-flipping contribution to conductance: the probability of the spin inversion 
from one $\sigma_z$ eigenstates to the other upon transition across the fluorinated segment.
(c) The amplitude of the scattering wave function for the peak at $E=0.022$ eV marked by the black dot in (b).
 } \label{noB}
\end{figure}

\subsection{Transport across a fluorinated layer at $B=0$}
The results of the preceeding subsection demonstrate that the spin flip is possible upon electron transition under many $F$ adatoms.
Nevertheless, an experimental fabrication of the extremely narrow ribbon with ordered $F$ positions  is rather  unlikely.  
Let us then consider the ribbon which is 292 atoms wide with random locations of the $F$ atoms [see Fig.~\ref{fig:292-map}(a)].

The calculated electron transfer probability as a function of the Fermi energy is given in Fig.~\ref{noB}(a). 
The blue line indicates the ribbon conductance in the absence of the fluorine atoms, which in $G_0$ units is equal to the number
of subbands that carry the flux to the right.
The adatoms perturb the system and induce strong backscattering, which induces the drop of conductance in Fig.~\ref{noB}(a) with
introduction of the adatoms. In the scattering region only 0.5\% of the carbon atoms are fluorinated, but the perturbation
of the potential by adatoms is strong and their random locations rule out the transparency of the system at resonances as in 
the ordered system of Fig.~\ref{trans}.

Only a very low spin-flip probability is observed [Fig.~\ref{noB}(b)].
A local maximum of the spin-flip probability near 0.022 eV [Fig.~\ref{noB}(b)] is associated with a quasi-bound resonance [Fig.~\ref{noB}(a)]
that is supported by a group of adatoms [see the amplitude of the scattering wave function at Fig.~\ref{noB}(c)]. 

The electron backscattering  by adatoms inverts the direction of the electron motion, and on the electron way back the spin precession \cite{besza} due
to the local Rashba interaction is reversed [see $\Delta \varphi$ for the incident and $-\Delta \varphi$ for the backscattered motion in Fig.~\ref{szc}(a)], hence the near cancellation of 
the overall spin precession events that results in the low spin-flipping conductance contribution in Fig.~\ref{noB}(b).

Figure \ref{noB}(b) shows that the spin-flipping effects of the electron passage across the fluorinated layer are very weak. 
Formation of a resonance supported by adatoms [Fig.~\ref{noB}(c)] with the Fermi electron experiencing a multiple
scattering  increases the electron dwelling time  in the area where the spin-orbit coupling is present and enhances
the spin-flip probability [see Fig.~\ref{noB}(b) near $E_F=0.022$ eV], which however remains very low.

\begin{figure}[htbp]
\centering
\includegraphics[width=0.30\textwidth]{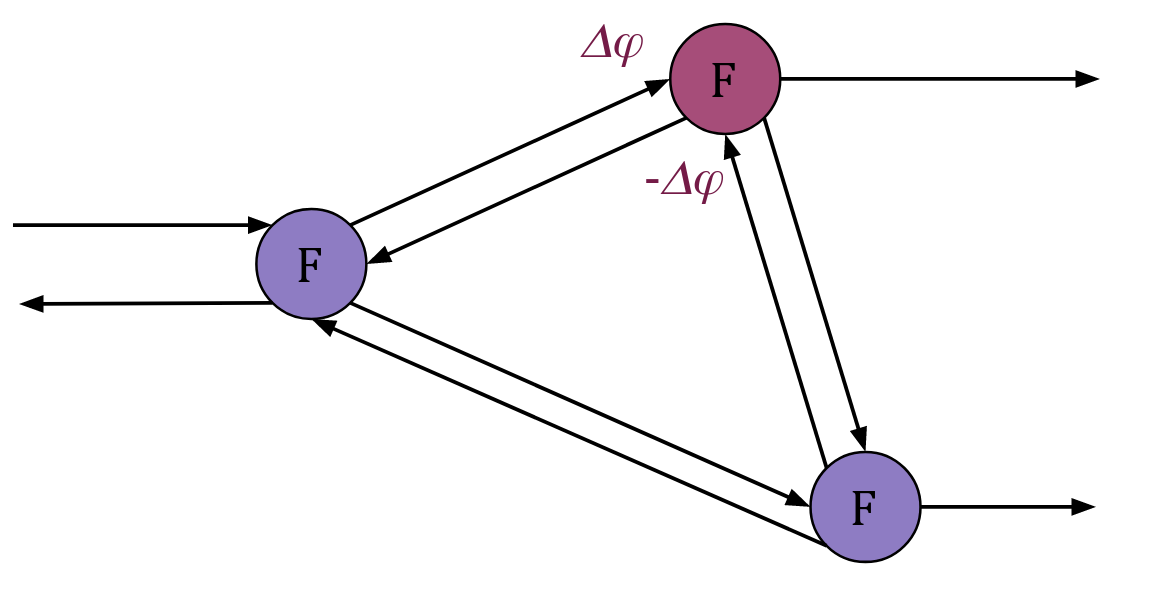}\put(-200,70){(a)}\\
\includegraphics[width=0.30\textwidth]{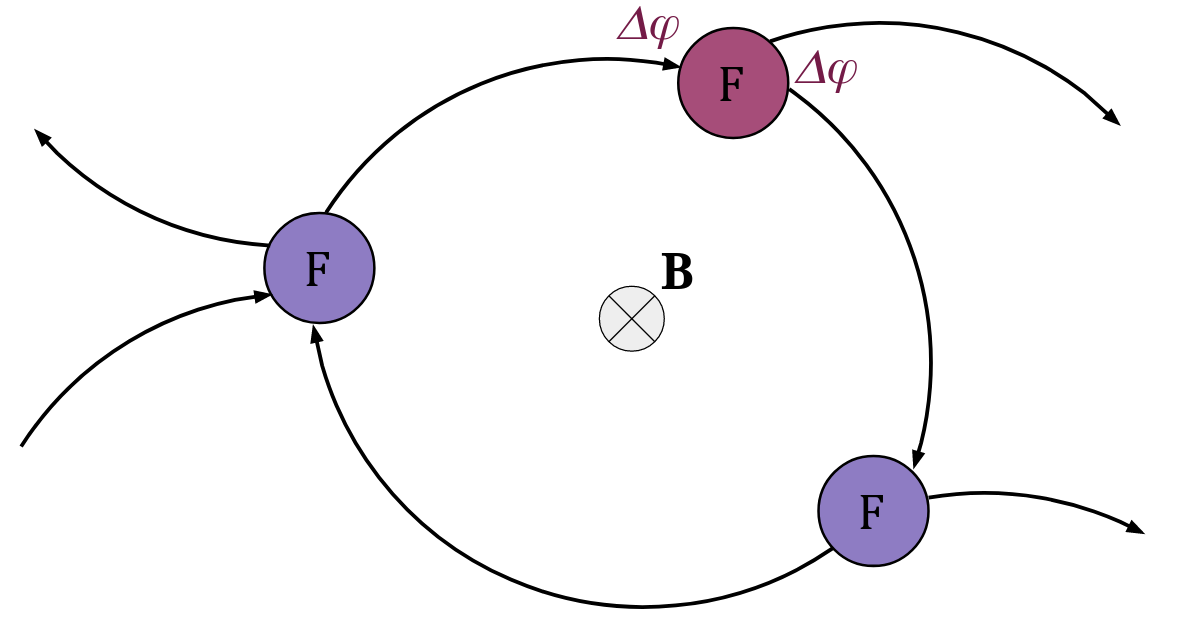}\put(-200,70){(b)}
\caption{Classical electron paths scattered by the fluorine adatoms (F). Electron is incident from the left. In the absence of the magnetic field (a) the electron gets backscattered
and the spin precession angle $\Delta \varphi$ due to the Rashba interaction induced by a single $F$ adatom marked by red is reversed by the backscattering $-\Delta\varphi$.
In the quantum Hall conditions (b) backscattering along the same path is absent due to  the magnetic deflection of the trajectories. For closed
paths the precession angle accumulates at each  electron transition near an adatom.} \label{szc}
\end{figure}

% UWAGA W HALL - BLOKOWANE BACKSCATTERING ALONG THE SAME PATH

%\begin{figure}[htbp]
%\centering
%\includegraphics[width=0.4\textwidth]{transz.pdf}
%\caption{The Aharonov-Bohm oscillations of conductance for a clean ribbon with the n-p junction induced as in Fig. \ref{s2} }\label{clean}
%\end{figure}

%\begin{figure}[htbp]
%\centering
%\includegraphics[width=0.49\textwidth]{psi-F-413.png}
%\caption{The  amplitude of the scattering wave function for the clean ribbon at a conductance peak at $B=41.3$ T of Fig. \ref{clean} (black dot).} \label{cleant}
%\end{figure}

\begin{figure}[htbp]
\centering
\includegraphics[width=0.49\textwidth]{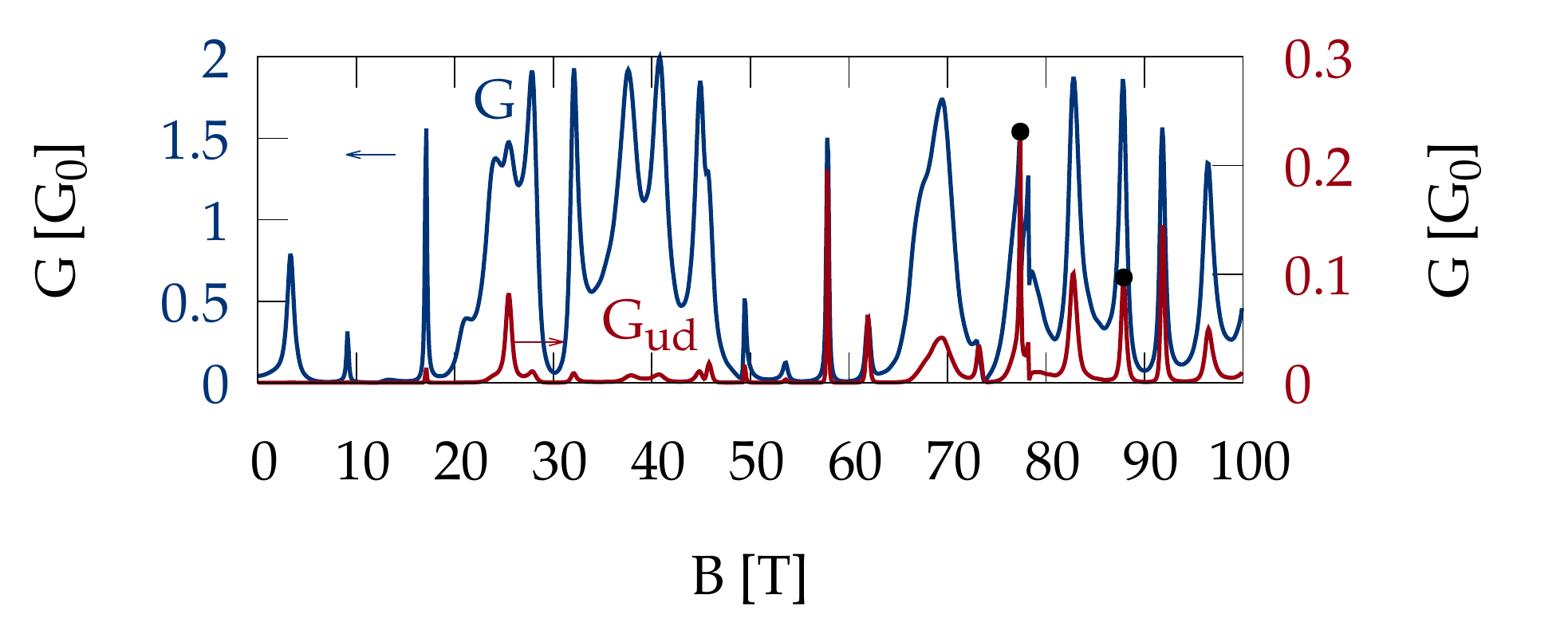}
%\includegraphics[width=0.49\textwidth]{data/wyniki/notip/trans.pdf}
%\put(-425,120){\textbf{292 (półprzewodnikowa) z tipem}}% lewa
%\put(-200,120){\textbf{292 (półprzewodnikowa) bez tipu}} %prawa
%\put(-445,100){(a)}
%\put(-220,100){(c)}
%\\
%\includegraphics[width=0.49\textwidth]{trans-spin.pdf}
%\includegraphics[width=0.49\textwidth]{data/wyniki/notip/trans-spin.pdf}
%\put(-445,100){(b)}
%\put(-220,100){(d)}
\caption{The overall (left axis, blue line) conductance and its spin flipping contribution (right axis, red line)
for the fluorinated n-p juncion. Black dots represent peaks for which scattering densities are presented below.  \label{xcx}}
\end{figure}

\begin{figure}[htbp]
\centering
\includegraphics[width=0.5\textwidth]{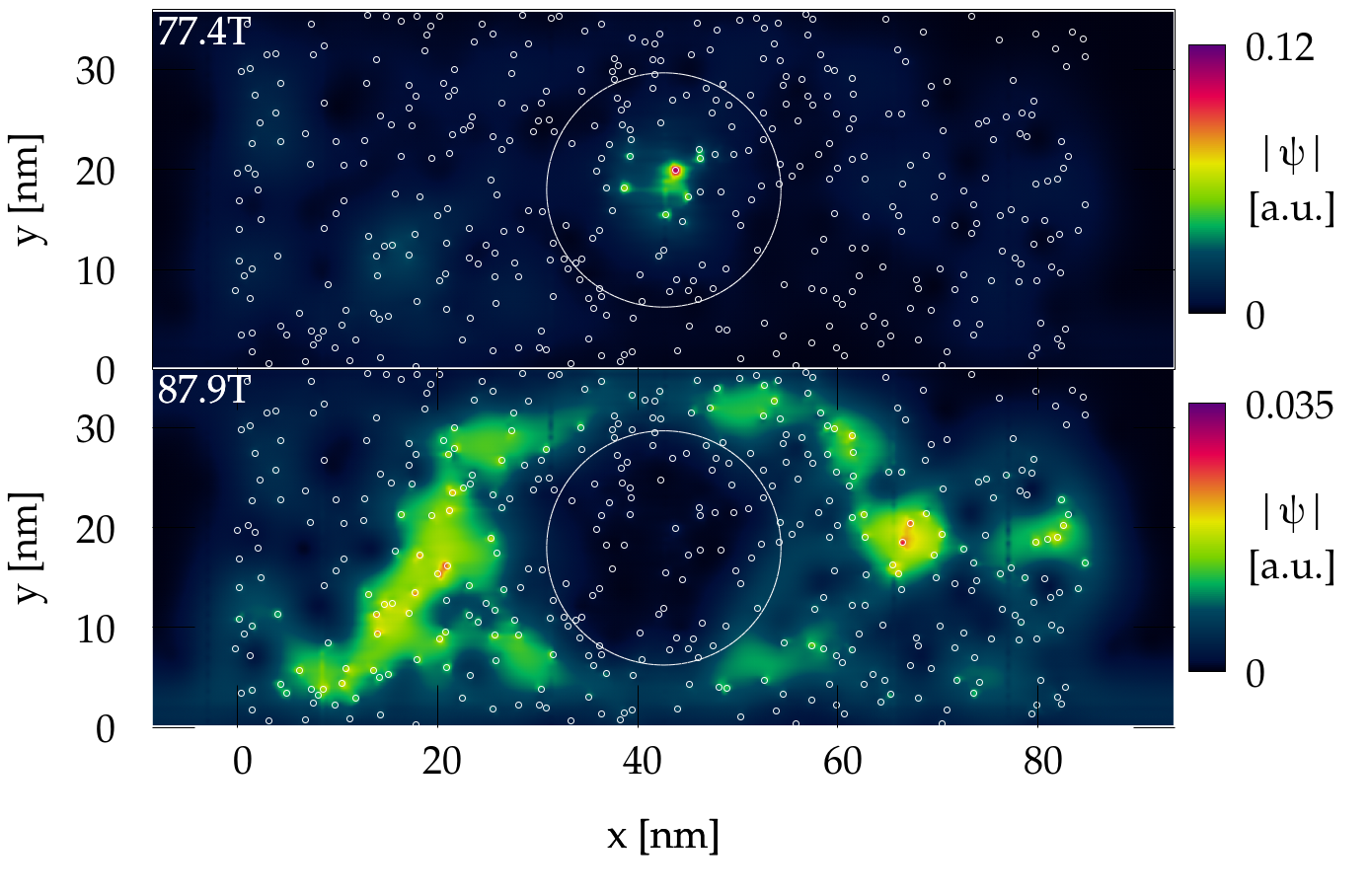}
\put(-230,100){ {\color{white}(a)}}
\put(-230,35){ {\color{white}(b)}}
\caption{The amplitude of the scattering wave function for two peaks of conductance of Fig.~\ref{xcx}  (black dots) at 77.4 T (a) and 87.9 T (b). The circle
indicates the position of the n-p junction defined by $E_F=V(x,y)$. The fluorine adatoms are marked with tiny white circles. }
\label{np}
\end{figure}

\begin{figure}[htbp]
\centering
\includegraphics[width=0.49\textwidth]{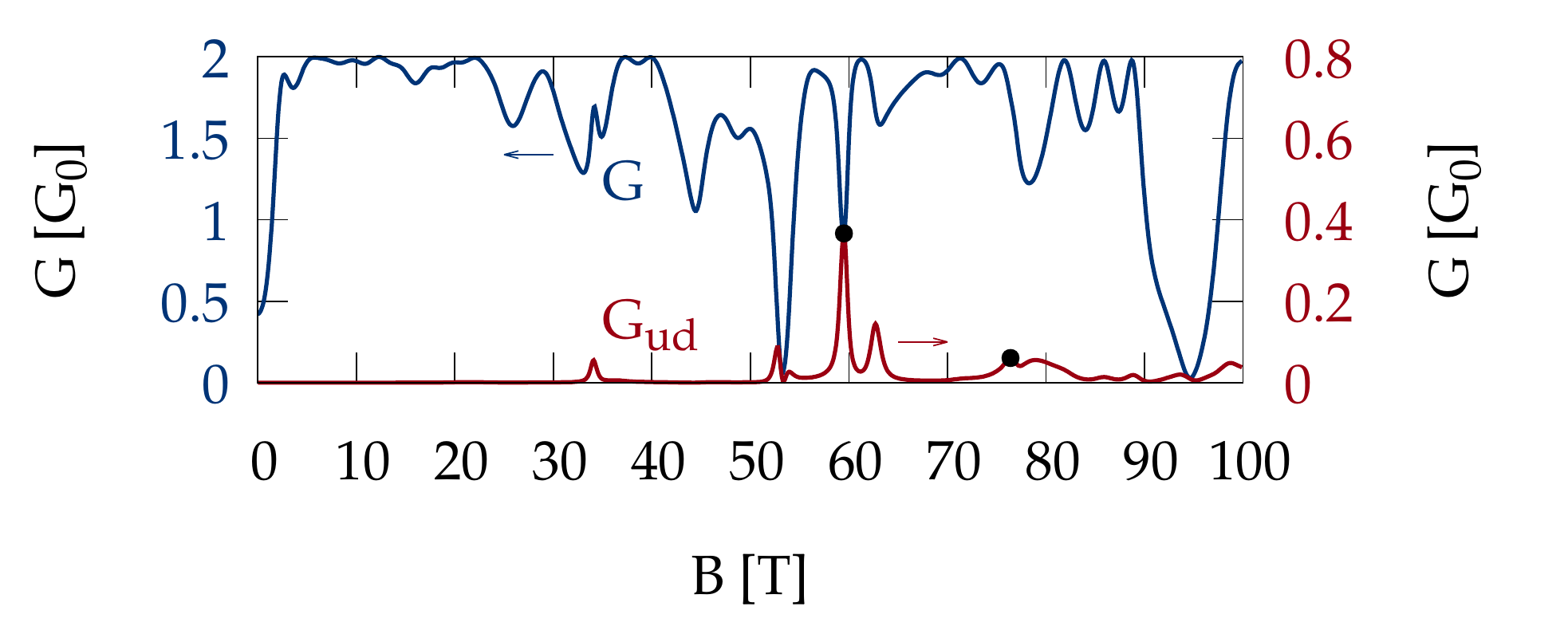}
%\put(-425,120){\textbf{292 (półprzewodnikowa) z tipem}}% lewa
%\put(-200,120){\textbf{292 (półprzewodnikowa) bez tipu}} %prawa
%\put(-445,100){(a)}
%\put(-220,100){(c)}
\\
%\includegraphics[width=0.49\textwidth]{data/wyniki/tip/trans-spin.pdf}
%\includegraphics[width=0.49\textwidth]{data/wyniki/notip/trans-spin.pdf}
%\put(-445,100){(b)}
%\put(-220,100){(d)}
\caption{Same as Fig.~\ref{xcx} only without the external potential defining the n-p junction.  \label{szasza}}
\end{figure}

\begin{figure}[htbp]
\centering
\includegraphics[width=0.49\textwidth,page=2]{psiFraq-NO-tip-292.pdf}
\put(-248,100){(a)}\\
\includegraphics[width=0.49\textwidth]{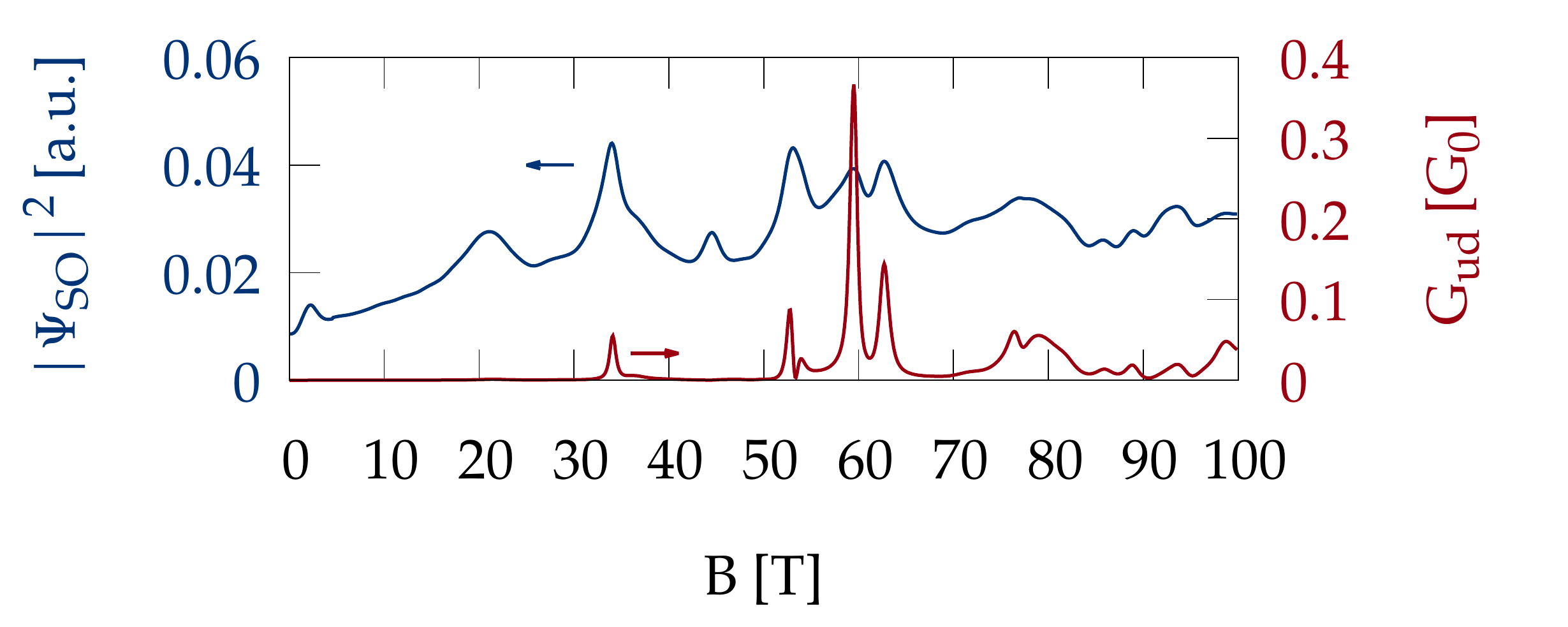}
\put(-248,100){(b)}\\
\includegraphics[width=0.49\textwidth]{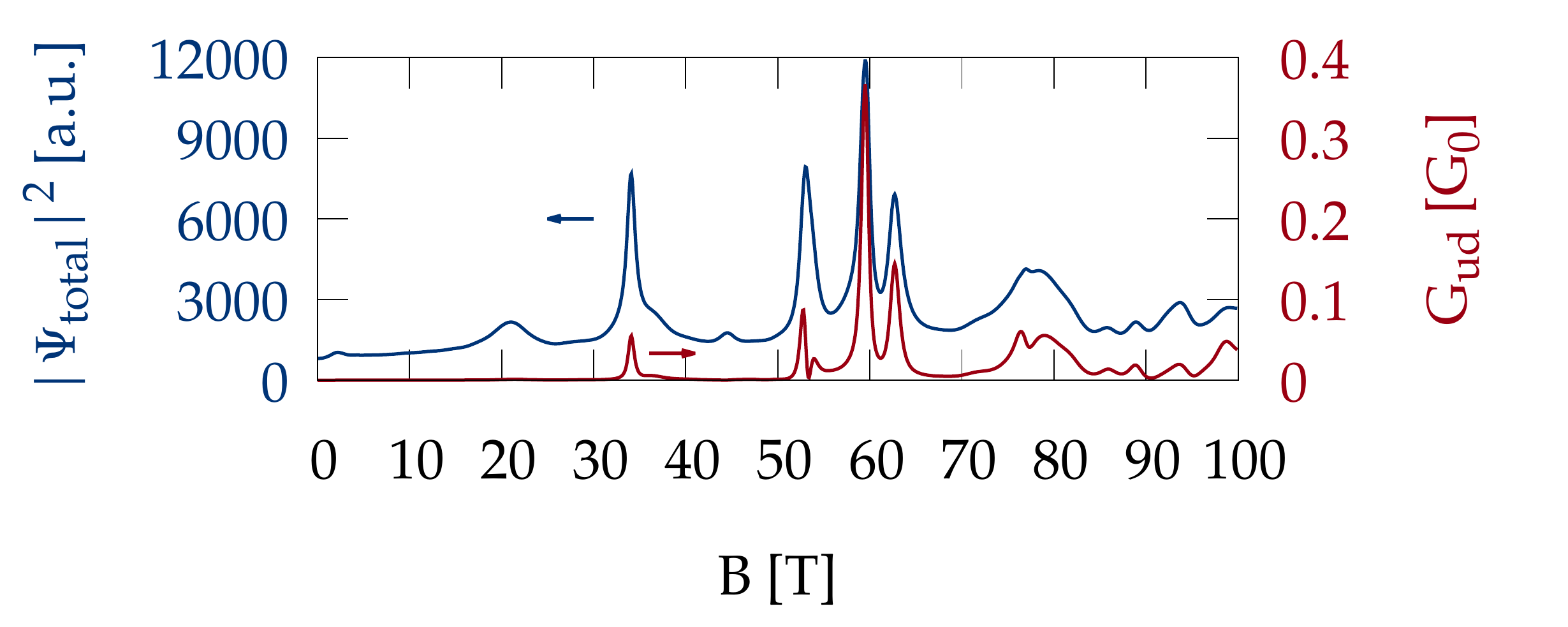}
\put(-248,100){(c)}\\
%\includegraphics[width=0.49\textwidth]{data/wyniki/psiFrak/psiFraq-NO-tip-292.pdf}
%\put(-445,130){\textbf{292 (półprzewodnikowa) z tipem}}% lewa
\caption{(a) The spin conserving component of conductance (orange line, right axis) and the fraction of the scattering density
localized in the area where the spin-orbit coupling interaction is present (blue line, left axis). (b) same as (a), only the red
line indicates the spin-flipping contribution to conductance. (c) same as (b) only with the blue line that indicates
the integral of the scattering density within the entire fluorinated ribbon segment for the normalized electron incidence amplitude.}\label{znormal}
\end{figure}

\begin{figure}[htbp]
\centering
\includegraphics[width=0.5\textwidth]{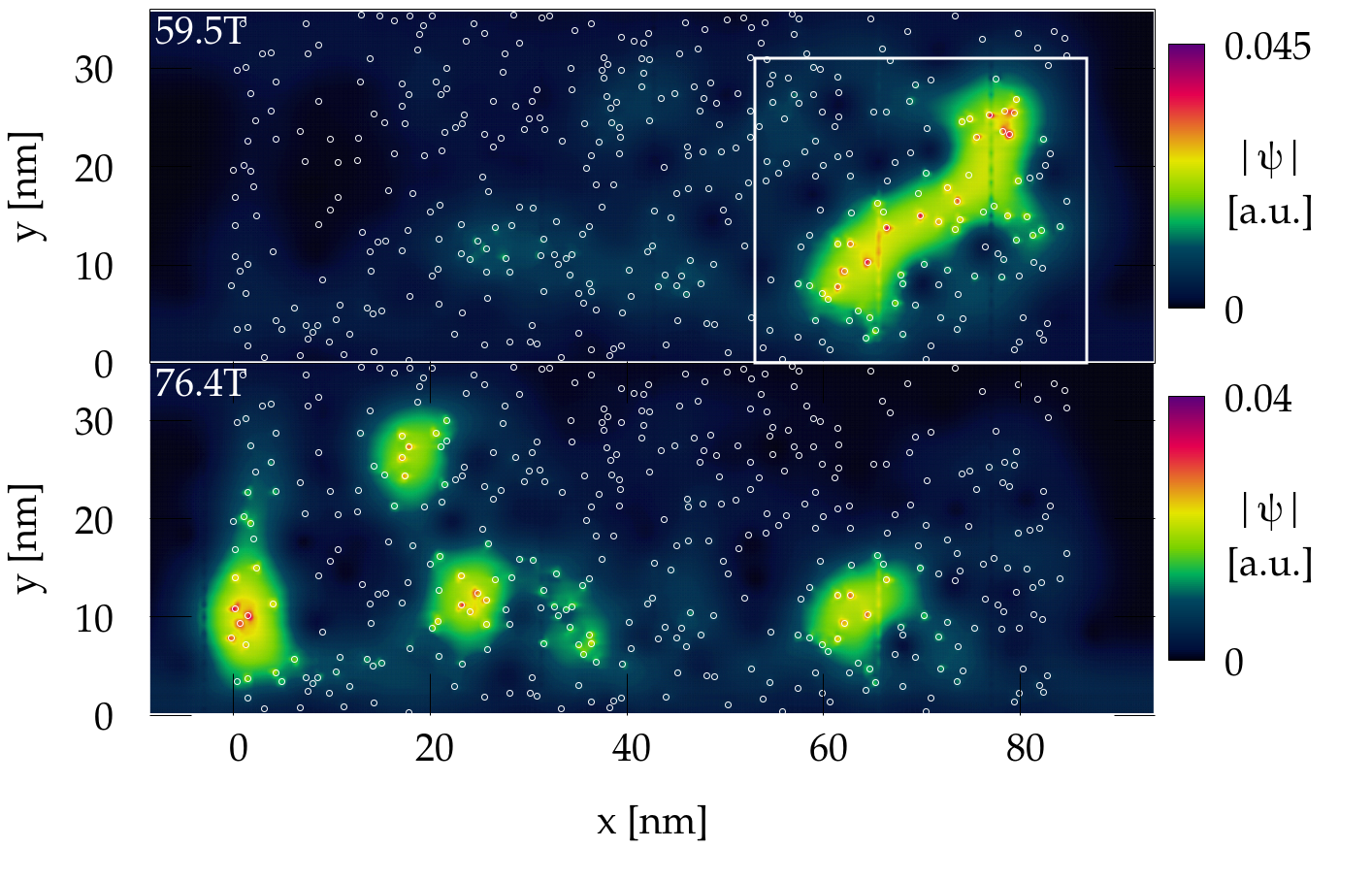}\put(-230,100){ {\color{white}(a)}}
\put(-230,35){ {\color{white}(b)}}

%\put(-320,385){\textbf{292 (półprzewodnikowa) z tipem}}% lewa
\caption{The amplitude of the scattering wave function for two peaks of conductance
of Fig.~\ref{szasza} (black dots) at 59.5 T (a) and 76.4 T (b). The
fluorine adatoms are marked with tiny white circles.}
\label{nonp}
\end{figure}

%\begin{figure}[htbp]
%\centering
%\includegraphics[width=0.4\textwidth]{data/wyniki/interf/zlacze.png}
%\caption{do poprawy}
%\end{figure}

\subsection{Recycling the electron passages: circulation around n-p junction (Zeeman interaction neglected)}

In order to accumulate the effects of the spin precession
we introduced a n-p junction to the system defined by a gate potential -- a tip of an atomic force microscope 
(see Fig.~\ref{s2}).
In the perpendicular magnetic field, it is possible to induce long-living resonances localized at the n-p junction
introduced by an external gate \cite{mre1}.  Moreover, the electron backscattering along the same path as in Fig.~\ref{szc}(a) is no longer
present in the quantum Hall conditions. 
The schematics of the current circulation in the quantum Hall conditions
is given in Fig.~\ref{s2}. For the magnetic field oriented 
to the graphene plane ${\bf B}=(0,0,-|B|)$ the classical Lorentz force pushes  the moving conduction band carriers to the right of their momentum.
In consequence the incident and transfered current of conduction band electrons flows along the lower edge, while the backscattering
 mediated by the circular n-p junction goes through the upper edge. For the considered magnetic field orientation the current circulation around the n-p junction is clockwise (see Fig.~\ref{s2}),
and the currents are stabilized for this single current orientation only \cite{mre2}. %Formation of the waveguide at the n-p junction requires formation of the quantum Hall conditions within the system. 
Formation of the waveguide at the n-p  junction separated from the edge of the ribbon requires formation of the quantum Hall conditions.
The separation of the edge and junction current occurs when a cyclotron radius fits between the edge and the junction.
For the cyclotron orbit given by the  magnetic length $l_B=\sqrt{\hbar/eB}=26\mathrm{[nm]}/\sqrt{B\mathrm{[T]}}$, the ribbon width $w$ and the diameter of the junction $d$, the condition reads  $2l_B<(w-d)/2$, which for the ribbon with 292 atoms considered here produces the condition $B>12$ T.

In this subsection we introduce external magnetic field with Hamiltonian $H'_B$. The spin Zeeman effect is introduced later.

%For a clean nanoribbon without the fluorine atoms the conductance exhibits [Fig. \ref{clean}] oscillations due
%to the Aharonov-Bohm effect along the ringlike waveguide [\onlinecite{mre1}]. 
%The scattering density for a maximal $G$ is displayed in Fig. \ref{cleant} with a 
%pronounced confinement of the density along the n-p junction. 

For the fluorine atoms present the conductance  undergoes oscillations [Fig.~\ref{xcx}] which are not as perfectly periodic
as for the clean ribbon \cite{mre1}, nevertheless an approximate periodicity at higher $B$ can be noticed.
The contribution to the conductance from the spin-flip transfer is -- at higher magnetic field -- large, reaching $0.2G_0$. Moreover,
the spin flips -- at higher $B$ -- become periodic and correlated with the conductance maxima [cf. Fig.~\ref{xcx}].
Note, that the maximal spin-flip transfer probability is increased by as much as three orders of magnitude with respect to Fig.~\ref{noB}(b).

Figure \ref{np} shows the scattering charge density
at the carbon atoms for 77.4T and 87.9 T - for which the spin-flipping contribution to conductance presents local maxima [Fig.~\ref{xcx}].
Figure \ref{np}(b) presents  a typical scattering density distribution for the resonances encircling the n-p junction perturbed by the adatoms.
In Fig.~\ref{np}(a) a localization of the resonance  at an adatom in the central -- p conductivity -- region can be seen. The resonances
localized inside potential barrier were discussed in Ref.~\onlinecite{mre1}. Their long lifetime results from the direction of
current circulation \cite{mre1} that the Lorentz force shifts  to the center of the barrier and  keeps the scattering density off the n-p junction. 

Note, that the peaks of the overall conductance [Fig.~\ref{xcx}(a)] and its spin-flipping contribution [Fig.~\ref{xcx}(b)] are correlated already for $B\geq 10$ T
and the spin-flipping peaks increase in amplitude for higher $B$. This results from an increasing magnetic confinement of the currents
near the edge and the junction which decreases  the coupling between the edge currents and the 
circular junction currents [Fig.~\ref{s2}] \cite{mre1,mre2}. The effect results in 
reduction of the coupling of the n-p junction to the edge. The lifetime of the resonances is increased along with the effects of the accumulation of the spin precession phase
shifts.

\subsection{Resonances in the disordered sample  (Zeeman interaction neglected)}

The conductance with the external potential of the precedent section removed is plotted in Fig.~\ref{szasza}.
We observe 
 an aperiodic dependence of conductance  as a function of the external magnetic field. A  conductance dip at $B=0$ [Fig.~\ref{znormal}(a)]
is characteristic to the weak localization as for a disordered conductor.  In graphene the weak localization dip is observed when 
the intervalley scattering is strong \cite{wlg}. In the present paper the role of the atomic scale defects that induce the intervalley scattering is played by the adatoms themselves. 
The intervalley scattering length can be estimated by the average distance between the $F$ atoms which is  $\simeq $ 1.5 nm for the dilute 0.5\% F concentration. In the present paper we consider ideal edges of the ribbons. The defects of the edge introduce additional 
intervalley scattering in addition to the adatoms. When the edge is defected  the peaks of conductance for nonzero $B$ change in position and the weak-localization dip varies
in depth but the $G(B)$ dependence is not qualitatively changed.

In perpendicular external magnetic field the resonances supported by the adatoms are also associated with current circulation
from one scattering event to the other, and the backscattering along the same path which limits the spin precession effects does not occur [Fig.~\ref{szc}(b)]
due to the Lorentz forces. 
This opens a chance for accumulation of the spin precession effects as for circular n-p junction.
% This is indeed the case 
 High peaks of the spin-flipping contribution to conductance are found [Fig.~\ref{znormal}(b)]
with irregular positions at the $B$ scale. The peaks of spin flipping conductance are now correlated with the dips of the spin-conserving conductance
in contrast to the results obtained for the circular n-p junction discussed above. 
The scattering density in the absence of the n-p junction exhibits localization of the quasi-bound states [Fig.~\ref{nonp}] varying 
between one resonance and the other. 

We searched for the relation between the form of the scattering density, the conductance and its spin-resolved contributions.
In Fig.~\ref{znormal}(a,b) the blue line shows the relative contribution of the scattering density at the fluorinated carbon atom
and its neighbors to the overall scattering density inside the computational box. 
In Fig.~\ref{znormal}(a) we can see that the transfer probability is minimal when the density
localization around the fluorinated carbon atoms is large. 
On the other hand, the spin-flip transfer probability  [Fig.~\ref{znormal}(b)] is maximal when the scattering density near the fluorinated atoms is large [Fig.~\ref{znormal}(b)]. 
The results are due to the fact that the  fluorine atoms are  both sources of backscattering and the spin flips. 
We found [Fig.~\ref{znormal}(c)]  a much closer correlation of the spin-flip transfer probability  
 with the ratio of the scattering density localized within the entire fluorinated region.
In the present approach the incident electron density is normalized and kept constant for any $B$.
The system of the adatoms for some values of the magnetic field supports a long living resonance
at the Fermi energy. In these conditions the scattering density within the fluorinated region 
acquires large values. The integral of the scattering density [Fig.~\ref{znormal}(c)] 
over the fluorinated region reproduces
very closely the shape the spin-flip transfer probability as a function of the magnetic field. 

\subsection{Wide ribbons and the Zeeman interaction}
Formation of the current confinement at the n-p junction or resonances supported by adatoms presented above
were due to the orbital effects of the external magnetic field that for the thin ribbon 292 atoms wide appeared only for the fields of the order of 50 T.  In order to shift the magnetic field scale to lower 
values the wider ribbon is needed. The conductance for the fluorinated ribbon of width 125 nm is plotted in Fig.~\ref{widenoz}(a), still without the Zeeman interaction. The spin-flips occur already for $B$ of the order of 10 T and are correlated with formation of localized resonances within the fluorinated ribbon segment (Fig.~\ref{widenoz}(b)).

\begin{figure}[htbp]
\centering
\includegraphics[width=0.5\textwidth]{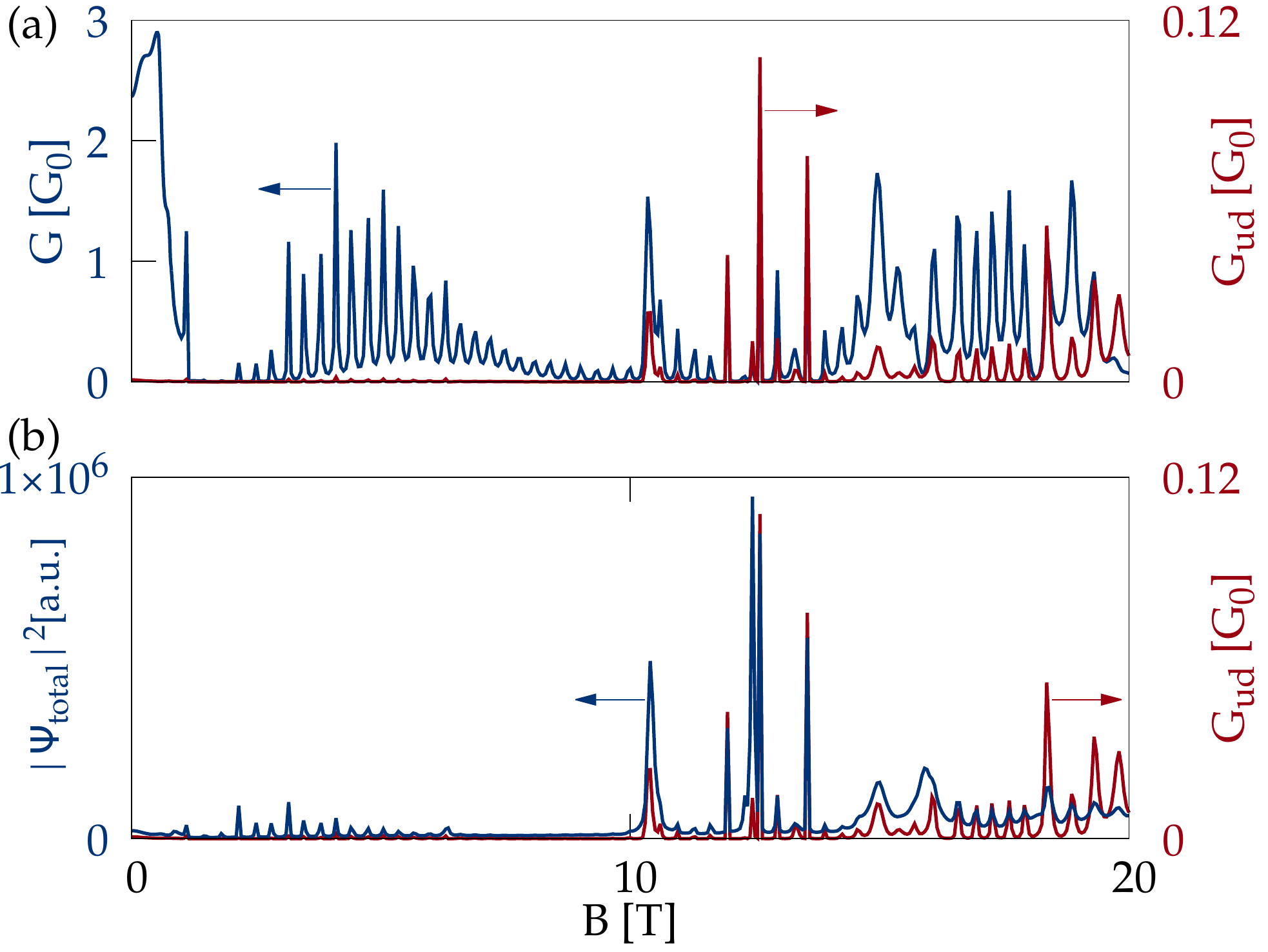}
%\put(-320,385){\textbf{292 (półprzewodnikowa) z tipem}}% lewa
\caption{(a) Conductance ($G$) and its spin-flipping contribution $G_{ud}$ 
for a wide nanoribbon with 1017 atoms across and width 125 nm for $E_F=30$ meV
(b) $G_{ud}$ versus the integral of the wave function within the fluorinated area.  Zeeman interaction is neglected.}
\label{widenoz}
\end{figure}

All the results presented above were obtained without the spin Zeeman effect. Figure \ref{zeeman} shows
the conductance as a function of the external magnetic field with the Zeeman interaction. 
The peaks of $G_{ud}$ appear but reduced by a factor of $\simeq 10$. Similarly, for the n-p junctions 
induced by external potential the spin-flipping contribution to conductance is drastically reduced by the spin Zeeman interaction. In presence of the spin Zeeman interaction for the perpendicular magnetic field the electron spin on its motion from one adatom to the other precesses with respect to the $z$ axis. This precession does not change the average $\langle S_z \rangle $ value, only  the perpendicular components $\langle S_x \rangle $ and $\langle S_y \rangle $ are affected. The spin precession length  at 10 T is 
1.785 \textmu m, i.e. by orders of magnitude larger than the average distance between the F-F adatoms ($\simeq 1.5$ nm). The effect behind the reduction of the  spin flipping conductance 
with the Zeeman interaction is the spin dependent Fermi wave vector. In the disordered sample the scattering wave function is very sensitive to the value of the Fermi wave vector and the accumulation of the spin precession events requires that the electron stays on its path while its spin is rotated. This is no longer the case when the spin Zeeman effect is present. The solution to this problem is given in the next subsection.

\begin{figure}[htbp]
\centering
\includegraphics[width=0.5\textwidth]{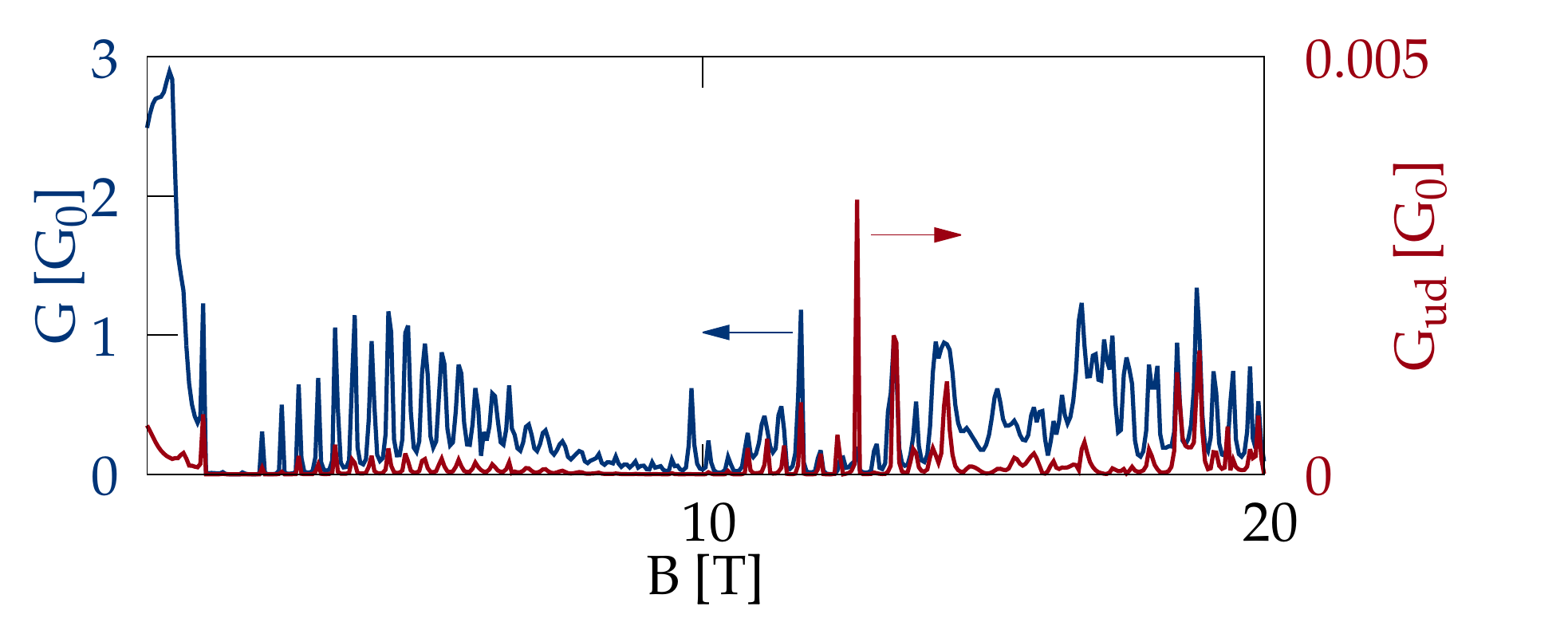}
%\put(-320,385){\textbf{292 (półprzewodnikowa) z tipem}}% lewa
\caption{Same as (a)  in presence of the Zeeman interaction. }
\label{zeeman}
\end{figure}

\subsection{Side-attached quantum rings} 
A way  to keep the electron on its path while the spin is rotated is the application of 
a lateral confinement of e.g. quantum ring as in Fig.~\ref{ring}. The quantum ring \cite{mre2}
supports localized resonances with the current circulation of a fixed orientation. 
The results for the ring attached to the thin ribbon with the Zeeman interaction is given 
in Fig.~\ref{ringg}. The spin-flipping contributions to conductance acquire values
in spite of the presence of the spin Zeeman interaction. 

 The overall conductance is a symmetric function of the external magnetic field
$G(B)=G(-B)$ in consistence with the Onsager relation for a two-terminal device.
Nevertheless, the spin-flipping contribution to conductance depends on the orientation of the magnetic field. The spin flips occur only for 
 $B<0$, i.e. for the magnetic field oriented in the $+z$ direction
which injects \cite{mre2,szmel} the incident electron wave function 
from the ribbon to the quantum ring. The injection occurs only  provided that a localized resonance is supported by the ring for the applied value of the magnetic field \cite{mre2}. 
 For strong magnetic field oriented in the $-z$ direction, the electron wave function is kept to the lower edge of the ribbon and does not notice the presence of the ring. This fact -- with the Onsager relation -- leads to $G(\pm B)=2G_0$ limit 
for high magnetic fields inducing the quantum Hall conditions (see Fig.~\ref{ringg}(a)).

\begin{figure}[htbp]
\centering
\includegraphics[width=0.5\textwidth]{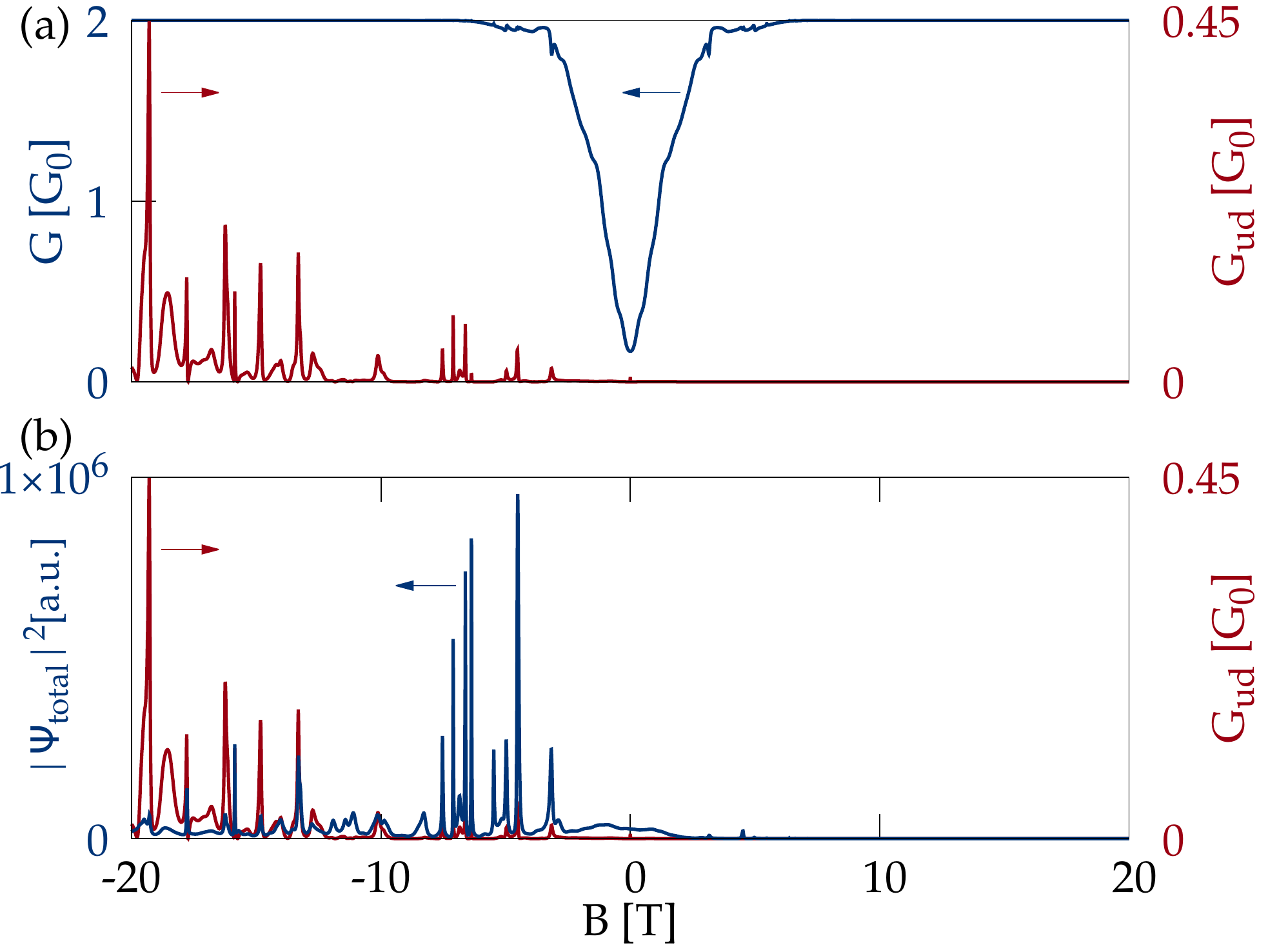}
%\put(-320,385){\textbf{292 (półprzewodnikowa) z tipem}}% lewa
\caption{ (a) The conductance $G$ and
  its spin-flipping contribution $G_{ud}$ for the fluorinated quantum ring of Fig.~\ref{ring} side attached
to the thin channel with 293 atoms across. (b) $G_{ud}$ versus the integral of the scattering wave function within the quantum ring. The spin  Zeeman interaction is present. The calculation is for $E_F=15$ meV and $\eta=1\%$ concentration of the fluorine atoms with tight binding parameters taken from \cite{irmer2015} for $7\times7$ supercell.}
\label{ringg}
\end{figure}

\section{Summary and conclusions}

We  studied charge and spin transport across a graphene nanoribbon with dilute fluorine adatoms
using the wave function matching technique within the tight-binding approach.
The electron passage below a single F adatom induces a small rotation of the electron spin 
due to the spin precession by a local Rashba interaction. In order to produce a spin-flip
many the micro precession events at separate adatoms need to accumulate. We demonstrated 
that the necessary accumulation occurs when the electron circulates around a closed path
in the external magnetic field in localized resonant states supported by an induced n-p junction
or by the adatoms themselves. The spin-flipping effect is deteriorated by the spin Zeeman interaction which 
introduces the spin dependence to the electron trajectory. The dependence is of a secondary importance
for a quantum ring side-attached to the nanoribbon which supports the localized resonances with a fixed
electron circulation around the ring and allows for large spin-flipping contribution to conductance
for the magnetic field orientation which injects the incident electron to the ring.

\section*{Acknowledgments}
This work was supported by the National Sci-
ence Centre (NCN) according to decision DEC-
2015/17/B/ST3/01161. The calculations were performed
on PL-Grid Infrastructure

%\begin{wrapfigure}{Hr}{0.5\textwidth}
%\centering
%\includegraphics[width=0.5\textwidth]{data/wyniki/tip/fluor-tip.png}
%\caption{Schemat. Fioletowe to fluor.}
%\end{wrapfigure}

\end{document}